\documentclass[%
 	reprint,
	showpacs,
	preprintnumbers,
 	amsmath,amssymb,
 	aps,
 	pra,
	]{revtex4-1}

	\usepackage{graphicx} 
	\usepackage{dcolumn} 
	\usepackage{bm} 
	\usepackage{natbib}
	\usepackage{hyperref}
		\hypersetup
		{
    			colorlinks=true,
    			linkcolor=blue,
    			anchorcolor=blue,
    			citecolor=blue,
    			urlcolor=black,
    			pdfborder={0 0 0},
    		}

\begin{document}

	\preprint{APS-PRA}

	\title{A New Stochastic Model of the Causal Interpretation of Quantum Theory on the Development of the Fundamental Concept of Mass}

	\author{Muhamad Darwis Umar}
 		\email{Darwis\_umar@ugm.ac.id}
		\affiliation{%
 		Department of Physics, Universitas Gadjah Mada, Sekip Utara BLS 21, 55281, Yogyakarta, Indonesia
	}%


	\date{\today}	

	\begin{abstract}

In this paper we pose two fundamental ideas on the motion of an elementary particle supporting the internal "spin motion" or \emph{Zitterbewegung} and a particle as concentrated energy. First, the particle moves randomly in a limited area (in a quantum-sized volume) like random vibrating system where the particle will diffuse in a quantum-sized volume when it absorbs or emits the quantized amount of energy. The quantum-sized volume can move too and plays role similar to the carrier amplitude, while the vibrational motion of the fundamental particle with frequency and amplitude represents a modulating signal. The current of diffusion process taking place in the quantum-sized volume will represent the emergence of a spin phenomenon shown by the existence of Clifford algebra. Second, the particle is pure energy concentrated on the surface of 3-dimensional sphere-form (2-manifold without boundary). Afterward we show that by defining the particle mass as an invariant quantity based on the two fundamental ideas, we can derive both the diffusion constant of the particle in the quantum-sized volume as $\beta = \hslash/2m$ and the Schr\"{o}dinger equation. Furthermore, by posing that the vibrational motion of the particle limited on the quantum-sized volume plays fundamental role as time interval unit (proper time$/$particle clock), we show that the relativistic effects of the particle must represent atomic process. 
		
	
	\end{abstract}

    \maketitle


	\section{\label{sec:1}INTRODUCTION}

Although there have been many attempts to interpret the causality principle of quantum theory \cite{IF-1946, IF-1952, WW-1954, DB-JPV-1954, EN-1966, AK-1985, RF-1933, KLC-ZZ-1985, DK-1964, KN-1986, MBJ-1988, THB-1970}, Nelson's work \cite{EN-1966} stands alone \cite{JRB-2002}. Nelson approaches the Schr\"{o}dinger equation in essence with stochastic mechanics as an open system, but he also used a non-friction model, thus there is no energy or mass transfers on average and the system can be kept as a closed system. Nonetheless, Grabert \textit{et. al} \cite{HG-PH-TT-1979} showed that quantum mechanics is not equivalent to a markovian diffusion processes, followed later by Gillaspie \cite{DTG-1994} who also demonstrated that the measurable behavior of most quantum system cannot be modeled as a Markov process. Skorobogatov and Svertilov \cite{GAS-SIS-1988} also pointed out that the measurable behavior of elementary quantum system can be modeled by non-markovian stochastic processes.

Stochastic interpretations, although they are based on the same fundamental idea on the existence of fluctuating field as background, have two major different viewpoints is considering the collapse of the wave function in the context of how we reconcile the probabilistic distribution of outcomes with deterministic form of Schr\"{o}dinger equation when the measurement is made. First, one views that when the measurement is made, the system will become an open system. This proposal serves the main idea of decoherence and de Broglie-Bohm interpretations on "measurement problem" is designated to keep the deterministic property of Schr\"{o}dinger equation. Second, one poses that the state collapse into macroscopically unique state \cite{GJ-PP-JR-2004} occurs dynamically because the wave function (internal mass density) is coupled to a Brownian motion noise term. This approach assumes that quantum mechanics is not exact from the beginning (before measurement is made). From a fundamental point of view, decoherence and de Broglie-Bohm interpretations have a problem in describing "reality" because it is just only one apparatus that exists in reality, while the dynamical collapse models do not explain the role of measurement well. \\
\indent Stochastic models with causality interpretations have been facing several challenges (questions) such as: 1. What is the origin of the non-classical force (related to osmotic velocity) acting on the particle or what is the origin of the quantum mechanical potential \cite{MPD-1979}?, quantity that related to internal motion \cite{ER-GS-1998}; 2. To derive stochastic models, one must introduce the diffusion constant by heuristic arguments, even though it has been shown that the Schr\"{o}dinger equation and diffusion equations are equivalent in a mathematical structure \cite{MN-1993}. The constant of diffusion expressed by Planck constant and mass has delivered a fundamental question: what is the natural significance of the Planck constant and its connection with mass concept in determining diffusion constant? 3. A realistic explanation of quantum mechanics has to meet at least one of the three requirements that is the charge of a particle is concentrated in a small volume of space \cite{KJ-2009}. 4. Nelson's model for stationery state contains the nodal surface caused by osmotic velocity that is inversely proportional to the probability density. 5. Stochastic models have not expressed yet an obvious picture of natural spin (a physical mechanism picture) that has been identified to play a fundamental role to determine the quantum behavior of micro-systems \cite{ER-GS-1998}. On the another word, stochastic models is not dealing with the existence of \emph{Zitterbewegung} that support proposal on splitting the motion of variables into two kinds of motion that are the motion of the center of mass to a chosen reference frame and internal motion with reference to the center of mass \cite{ER-GS-1998}.

Besides the measurement problem in the foundation of quantum theory, mass, behind its role as the most fundamental notion underlying physics to perceive and to conceptualize relations among all physical phenomena, has not yet been fully understood and explained. Fundamentally, the definition of mass is not only present in classical-relativity with terms inertial mass and relativistic mass as well as gravitation mass, but also found in quantum theory in its connection with the density of probability or the wave function. The unclear ontology role of mass to unify our understanding on what truly goes on in physical realm indicates that the concepts of mass are simply not complete. It means that definition of mass still open to be redefined and extended for unified purposes of consistently creating a natural picture. The definition, of course, should be derived from a new theory with which we can explain the origin, the existence and the phenomenological properties of mass, as well as the fundamental source of unpredictability and the measurement problem in quantum theory at once.

In this work, we pose a unified description on both the fundamental motion of particles and particle masses in the new stochastic process framework in order to produce a natural explanation of spin. From the proposal, we describe what happens to the particle when a measurement is made, what the relation between the deterministic evolution of the wave function and the probabilistic outcome from measurement is, and as well as what the connectivity between the particle and what its environment in the concentrated-energy context is. We also introduce a new meaning of special relativity theory in dynamics viewpoint and its connection with internal atomic mechanism.

	\section{\label{sec:2}THE FUNDAMENTAL MOTION OF A PARTICLE}

The debate on the interpretation of quantum mechanics has taken place since the 1920s and continues to this day in both the scientific and philosophical communities. One of issues that has become central to the debate is the possible presence of the role that causality plays in microscopic world governed by the laws of quantum mechanics \cite{PJR-2009}. Orthodox quantum theory known as the Copenhagen interpretation, which is the prevailing theory of quantum mechanics, confronts with the causality principle based on the notion of cause and effect. They, anti-realism, have a notion that in quantum realm, particles do not acquire some of their characteristics until they are observed by someone \cite{PJR-2009}.

Contrary to orthodox quantum theory, attempts to understand quantum theory in the classic understanding with the causality principle still continue in varying models. The original stochastic model (a non-local theory) was first introduced by Bohm and Vigier \cite{DB-JPV-1954} describing the random motion of the particle in Madelung fluid. Afterward, Nelson \cite{EN-1966} attempted to discuss Schr\"{o}dinger equation in the universal Brownian motion framework (a local theory). Stochastic interpretation by Nelson \cite{EN-1966} claimed that the quantum mechanical motion of particles governed by the Schr\"{o}dinger equation can be equally understood as particles in classical Brownian motion in a vacuum which acts upon the particle as does heat in the theory of irreversible processes \cite{HG-PH-TT-1979}. Nelson's approach is different from those of F\'{e}nyes \cite{IF-1946, IF-1952} and F\"{u}rth \cite{RF-1933} in that it uses an imaginary diffusion coefficient. Contrary to the notion of Bohm-Vigier and Nelson that viewing the wave function evolves in deterministic way, Ghirardi, \textit{et. al.} \cite{GCG-RA-TW-1986, GCG-PP-AR-1990} posed that the wave function interacts with background field fluctuations, therefore the probability outcome should include background field fluctuations. Actually, the proposal of Ghirardi, \textit{et. al.} \cite{GCG-RA-TW-1986, GCG-PP-AR-1990} (known as GRW and CLS models) are not interpretation models of quantum theory, but rather, models to understand measurement problem.

If we trace stochastic models, we will find out that almost all them use closely similar causal interpretations, which are
		\begin{enumerate}
			\item The change in trajectory of an object (a particle) is continuous and that in instant of time it has some definite positions \cite{DB-1989}.
			\item Vacuum fluctuation plays a central role in determining diffusion and random processes.
		\end{enumerate}

Here we introduce a new fundamental mechanism of the motion of a particle that is entirely different from the previous models. Although we build our model in terms of random motion (the change of motion trajectories), fundamentally, we have radically departed from traditional views. We pose that random motion in a quantum system is an intrinsic property of particles, and this motion is naturally caused by concentrated energy localized as a particle. It means that localized energy will play a role as quantum potential. Random motion takes place in a quantum-sized volume (it is similar to random vibrating motion) where this volume can move too, and then mass and charge distributed in the quantum-sized volume equal to the distribution of probability to find the particle in the quantum-sized volume.

There are four mechanism introduced in our model, which are as follows.
		\begin{enumerate}
			\item The particles will move randomly in a quantum-sized volume where the random motion serves an intrinsic property of particle and related to spin phenomena. In a stationer state, random motion will relatively take place forward and backward in time with markovian process. Nevertheless, because random motion takes place in limited area (the quantum-sized volume) that similar to random vibrating system, random motion does not present perspective on the diffusion process.
			
			\item Relative to random intrinsic motion, the quantum-sized volume can undergo translational movement, but it cannot take place backward in time at stationery state. The dynamics of the quantum-sized volume represent magnetic properties of angular or translational momentum, and it can represent de Broglie-Bohm theory on guiding particle.
			
			\item Every change of the speed of random intrinsic motion accompanied or not by the change of the velocity of translational motion of a quantum-sized volume will represent the transition between two states, and every change of the speed of random intrinsic motion will generate a diffusion process or Brownian motion perspective. The dissipative character of quantum systems in our model is similar to Langevin's approach to Brownian motion but with a different mechanism. The difference lies in the cause of emission or absorption process. In our model, a particle will emit or absorb energy when the intrinsic speed of the particle and or the translational velocity of quantum-sized volume changes, and it is not related to a friction process. Internal diffusion process coincide with emission and absorption processes which also occurs in backward and forward in time with a non-markovian process (as long as the system does not undergo phase transition).
			
			\item For more any complex physical system (material), the quantum-sized volume can probably move in two ways that are transitional or random motions, and it will be determined by the potential forms of system (material).
		\end{enumerate}

If we consider the kinematic aspects of the transition process (how the particle moves from one point to other point, i.e. how its position changes in time); for example, a case where transition processes include diffusion process and the movement of the quantum-sized volume, then the movement path of the particle at any time will be determined by the velocity of the quantum-sized volume, the distribution of the velocity field in the quantum-sized volume describing how the particle relatively moves from the center of quantum-sized volume, and the displacement due to the diffusion process. We can represent the forward time operator for describing how the average position of the particle changes in time when the transition process followed by emitting energy occurs (the speed of the intrinsic motion in the quantum-sized volume decreases) as \cite{EN-1966}:	
		\begin{align}
			\label{1}
			D = \frac{\partial}{\partial t} + \mathbf{b} \cdot \nabla + \beta \nabla^2.	
		\end{align}		
While the backward time operator for the case of absorbing energy followed by the increase of the speed of the quantum-sized volume is
		\begin{align}
			\label{2}
			D^{~}_* = \frac{\partial}{\partial t} + \mathbf{b}^{~}_* \cdot \nabla - \beta \nabla^2,		
		\end{align}	
where $\beta$ is the diffusion constant. If we consider a transition process followed by emitting energy. then $\mathbf{b}$ is the initial velocity field of a stationer state and $\mathbf{b}^{~}_*$ is a relatively final velocity where $\left|\mathbf{b}| < |\mathbf{b}^{~}_*\right| $.

Next we consider Langevin's equation describing energy emission when a particle undergoes a transition process generated by applying an external force (an external potential). We assume if the physical system does not undergo emission or absorption processes, then every particle of any physical system is in stationary states. If an external force (an external field) is applied to any physical system being in any stationary state where energy of the external field is equal to the gap energy between stationary states so that the interaction will produce emission or absorption processes (between) as long as the interaction don't cause any phase transition. In such process, the set of possible stationary states of the physical system do not change by external field. But, if any applied external field do not generate transition process (diffusion processes: emission or absorption processes), then the physical system will respond to the presence of the external field by creating a new stationary state characterized by changing its particle positions and kinetics so that all new possible states produced by the external potential are different from previous possible stationary states (without any external field). Our starting point to consider any physical system at the stationery state is based on the historical facts that quantum theory was develop to understand the behaviors and the properties of atom and molecule as well as other stationer physical systems. We apply Langevin's equation
		\begin{align}
			\label{3}
			m\, \mathbf{a}^{~}_{\mathrm{net}}
				&= \mathbf{F}^{~}_{\mathrm{qv}}\bm{(}(\mathbf{x}(t)\bm{)} 
					- \xi \mathbf{b} \bm{(}\mathbf{F}^{~}_{\mathrm{external}}(t)\bm{)} \nonumber \\ 
				&\qquad	
					+ \mathbf{F}^{~}_{\mathrm{external}}(t) + \mathbf{F}^{~}_{\mathrm{random\,force}}(t),
		\end{align}
where $\mathbf{x}(t) = \mathbf{R}(t) + \mathbf{r}(t)$, $\xi$ is the friction coefficient, and $\mathbf{a}^{~}_{\mathrm{net}}(t) = [\ddot{\mathbf{R}}(t) + \ddot{\mathbf{r}}(t)]$ is the total acceleration where $\ddot{\mathbf{R}}(t)$ is the acceleration of the quantum-sized volume and $\ddot{\mathbf{r}}(t)$ is the acceleration of particle in the quantum-sized volume relative to the center of volume, $\mathbf{F}_{\mathrm{qv}}\bm{(}\mathbf{x}(t)\bm{)}$ is the force governing the stationer state of the quantum-sized volume [we can also consider an ideal condition in which $\mathbf{F}^{~}_{\mathrm{qv}}\bm{(}\mathbf{x}(t)\bm{)} = 0$], $\mathbf{F}^{~}_{\mathrm{external}}(t)$ is the applied force for generating transition process, and $\mathbf{F}^{~}_{\mathrm{random\,force}}(t)$ is random force that in the conventional view is asserted to represent the effect of background noise, but in our model, by contrast, we pose this force only represent an intrinsic internal random motion.

We can rewrite Eq. (\ref{3}) in terms of velocity as
		\begin{align}
			\label{4}
			\mathrm{d}\mathbf{v}^{~}_{\mathrm{net}}
				&= -\frac{1}{m}\nabla V\bm{(}\mathbf{x}(t)\bm{)}\,\mathrm{d}t 
					- \frac{\xi}{m}\mathbf{b}\bm{(}\mathbf{F}^{~}_{\mathrm{external}}(t) \bm{)} \,\mathrm{d}t \nonumber \\
				&\qquad + \frac{1}{m}\mathbf{F}^{~}_{\mathrm{external}}(t)\,\mathrm{d}t + \frac{1}{m}\mathbf{F}^{~}_{\mathrm{random\,force}}(t)\,\mathrm{d}t,
		\end{align}	
where $\mathbf{v}^{~}_{\mathrm{net}}(t) = \dot{\mathbf{R}}(t) + \dot{\mathbf{r}}(t)$. Applying average forward time derivative to Eq. (\ref{3}), we find that
		\begin{align}
			\label{5}
			D\mathbf{v}^{~}_{\mathrm{net}}
				&= -\frac{1}{m}\nabla V\bm{(}\mathbf{x}(t)\bm{)} 
					- \frac{\xi}{m}\mathbf{b}\bm{(}\mathbf{F}^{~}_{\mathrm{external}}(t)\bm{)}\nonumber \\
				&\qquad +\frac{1}{m}\mathbf{F}^{~}_{\mathrm{external}}(t).
		\end{align}				
Whereas for average back forward time derivative to Eq. (\ref{3}), we find		
		\begin{align}
			\label{6}
			D^{~}_*\mathbf{v}_{\mathrm{net}}
				&= -\frac{1}{m}\nabla V\bm{(}\mathbf{x}(t)\bm{)} 
					+ \frac{\xi}{m}\mathbf{b}\bm{(}\mathbf{F}^{~}_{\mathrm{external}}(t)\bm{)} \nonumber \\
				&\qquad + \frac{1}{m}\mathbf{F}^{~}_{\mathrm{external}}(t).
		\end{align}
$\mathbf{b}\bm{(}\mathbf{F}^{~}_{\mathrm{external}}(t)\bm{)}$ in Eqs. (\ref{5}) and (\ref{6}) shows that diffusion processes representing emission and absorption processes are caused by $\mathbf{F}^{~}_{\mathrm{external}}(t)$, Thus, the absence of $\mathbf{F}^{~}_{\mathrm{external}}(t)$ represents the physical system is in stationary state.

Seeing Eq. (\ref{5}) and (\ref{6}), it seems that the two equations describe ambiguous mechanisms. It is because the two equations use the same forces to describe two difference mechanisms. We can understand it with following description. Eq. (\ref{5}) depicts external force applied on a particle occupying a stationary state with potential $V\bm{(}\mathbf{x}(t)\bm{)}$ and then makes the particle undergo a friction force (emission process). Basically, this process will create the change of internal potential where the change of internal potential will be the same as the external force. Thus we can still consider Eq. (\ref{6}) as a backward process with the same internal potential and the same external force, but it takes place spontaneously (without external treatment). From this point of view, we can view the spontaneously emitting or absorbing processes as backward processes as long as the system does not undergo a phase transition. Since backward processes play a role as a natural tendency to bring final states (as the results of forward process) back to initial states (before there are external treatments) so that we pose that there are neither spontaneously emitting or absorbing processes without interaction between any physical system and external treatment or perturbation. Thus the information of any dynamic system must only be acquired and accessed by applying external treatment to any physical systems occupying stationary states. Using the mean forward derivative (\ref{1}) and the mean backward derivative (\ref{2}) introduced by \cite{EN-1966}, Eqs. (\ref{5}) and (\ref{6}) give
		\begin{align}
			\label{7}
			\frac{1}{2}\left(D D^{~}_* + D^{~}_* D\right)\mathbf{x}(t)
				&= \mathbf{a}^{~}_{\mathrm{net}}(t)\nonumber \\
				&= -\frac{1}{m}\nabla V\bm{(}\mathbf{x}(t)\bm{)} + \frac{1}{m}\mathbf{F}^{~}_{\mathrm{external}}(t),
		\end{align}				
where $\mathbf{x}(t) = \mathbf{R}(t) + \mathbf{r}(t)$. 

For stationer state $\left[\mathbf{F}_{\mathrm{external}}(t) = 0\right.$ and $\left.\pm v \mathbf{b}(t) = 0\right]$, Eq. (\ref{7}) becomes
		\begin{align}
			\label{8}
			\frac{1}{2} \left( D D^{~}_* + D^{~}_* D\right)\mathbf{x}(t)
				= \mathbf{a}_{\mathrm{net}}(t)
				= -\frac{1}{m}\nabla V\bm{(}\mathbf{x}(t)\bm{)}.
		\end{align}

	\section{\label{sec:3}THE NEW FUNDAMENTAL CONCEPT OF MASS}

If we assume that an obvious connection between the microscopic (quantum realm) and macroscopic worlds (classical picture) exists, then every proposed theory must not only reveal the unfinished understanding of how quantum realm is more obscure than our daily imagination and  physical sense that perceive physical phenomena based on both conventional models about particle, interactions and concepts of mass, charge, field and energy in which all those physical concepts and models constitute physical pictures via mechanisms obeying causality principle. But, the proposal must also solve the polemic about whether mass and charge are no more than the abstract quantitative expression of facts that do not need to 'explain' phenomena in terms of purposes or hidden causes like Weyl's and Mach opinion on mass definition \cite{MJ-2000}. In our opinion, the existence of a hidden connection between the quantum and the classical realms may be caused by the fundamentally unenviable approach of describing the fundamental attributes of physical objects that have been causing and afterward shaping our incomplete perspective in understanding all physical phenomena and relations among them such as mass, time, force, field, interaction, etc.

On of the fundamental concepts that has not been established yet is the concept of mass. How modern physics has both experimentally and theoretically contributed to a more profound understanding of the nature of mass has been comprehensively summarized and reported by Max Jammer in Ref. \cite{MJ-2000}. Mass was formally introduced by Newton in his second law of dynamic to depict the dynamics of physical object via the relation of force and acceleration. This description has provided the inertial feature of physical objects. Newton also presented the gravity mass to describe the ability of matter to generate the phenomenon of gravity force. Afterward, the meaning of mass was amended by Einstein when he introduced special relativity theory extends our understanding of mass through the differentiation of rest mass and relativistic mass. Nevertheless the debate on the rest mass versus relativistic mass still exists until now \cite{MJ-2000}. An important aspect of Einstein's work has been the relation of energy and mass which leads to the perspective of a particle's mass at the rest as concentrated energy. Einstein's contribution to the perspective of mass also included a unified description of inertial mass and gravity mass when he worked on general relativity. In this context, mass is extended as a key factor governing the structure of space-time. Although Einstein had created a fundamental picture of the particle as concentrated energy using the term of the speed of light (the relation between mass and energy), the results cannot link the description of mass-energy with the origin of the electromagnetic field or the charge of the particle, another important physical concept which provides a perspective of the electromagnetic wave and its energy. The kinematic approach introduced by Einstein has not yet provided a complete description on what the meaning of electric and magnetic fields in relation to the mass feature of the charge particle is. Another missing explanation has been what the picture of the physical mechanisms on how electrical fields transform to magnetic fields or conversely when a charged particle moves (following the perspective of kinetic energy between two frames of reference)?; what is the effect of radiation (the production of photon) in regards to the mass of charged particle when the particle is accelerated?; how clock and length are connected with the atom and many body problems \cite{HRB-OP-2001}; until how energy is stored as bounding state and how it is released as electromagnetic field or photon. We view that non-unified description of the electromagnetic and mechanical aspects of a particle is the source of an unfinished-description on electromagnetic interaction and also the source of a debate on what is the physically invariant quantity or not?

Here, we pose a new description of mass that is a totally different from the conventional approaches. First, we give two backgrounds to our proposal.
		\begin{enumerate}
			\item From our model on the fundamental motion of a particle, we can identify that random motion in the quantum-sized volume is the intrinsic property of the particle, hence the terminology of energy (quantum potential) generating intrinsic motion should be connected to concentrated energy "rest mass" as in the special relativity theory perspective.
			
			\item Because the concentrated energy of the particle must decrease and increase by emission and absorption processes by external treatment, the amount of quantified-energy emitted or absorbed must relatively be representative of the structure and motion of the particle due to the existence of system (other particles/environment). This also means that energy related to transition processes will represent how other particles as an environment relatively evolves and changes with time towards the particle as a frame of reference. Conversely, transition processes will describe how the particle relatively evolves and changes with time toward their environment (how their potential and kinetic terms relatively change toward the other particle).
		\end{enumerate}				

To construct a new model based on the two possibilities above, we pose that:
		\begin{enumerate}
			\item The elementary particle is 3-dimensional sphere-form (2-manifold without boundary) where natural energy will occupy at the surface (a concentrated-energy system).
			
			\item When there is no perspective about system (environment) 'particle in absolute vacuum', energy occupying the surface of the particle makes the particle move randomly and rotate about its axis (spin motion) where the speeds of both translational random motion and internal angular motion is at the speed of light. The two kinds of fundamental particle motions take place in a quantum-sized volume as our previous model on fundamental of the particle.
			
			\item Random motion taking place in the quantum-sized volume allows the motion of the particle to be considered as a randomly vibrating-movement.
			
			\item We defined mass as the amount of energy per unit of surface area of the particle per its vibration and rotating frequencies, and the value of the mass of the particle is always constant for every state:
				\begin{align*}
					m = \frac{E}{4\pi \left(\bm{\mathfrak{R}}^2_{\mathrm{particle}}\right) 						
							\nu^{~}_{\mathrm{vib}}\nu^{~}_{\mathrm{rot}}} = \frac{E}{\mathbf{c}^2},
				\end{align*}
where $\mathbf{c}$ is light velocity field, $m$ is mass tensor (in any time interval), $\nu_{\mathrm{vib}}$ and $\nu_{\mathrm{rot}}$ respectively are vibration and rotational frequencies, $4\pi\left(\bm{\mathfrak{R}}^2_{\mathrm{particle}}\right)$ is the surface of the fundamental particle, and $\bm{\mathfrak{R}}$ is a vector field where its scalar represents the radius of sphere and its direction of random vibrating motion. If we pose that $E$ and $m$ are invariant to all frame of references so that $\mathbf{c}^2$ will be invariant quantity caused by the invariance of mass quantity.

			\item The concentrated energy of the particle at every state, for every physical system, must always be at a certain value that is less than that of its value when there is no interaction perspective (a particle is in absolute vacuum), whereas the loss of the concentrated energy of the particle will be transformed into the presence of any physical system as expressed by potential terms and kinetics terms and the changes of the speed of internal random motion to lower than the speed of light where the speed of the random motion in whatever kinds of physical systems must always be lower than the speed of light. Thus, interactions and transition states will only change the potential terms and or kinetics terms and or the speed of random motion. For example, the equation of the invariant mass of an atom consisting of $n$ electrons undergoing a transition process can be expressed by Eq. (\ref{9}) and {\ref{10}}: Consider an electron of the atom is at a stationery state; so invariant mass principle requires
				\begin{widetext}
				\begin{align}
					\label{9}
					m	= \frac{E - V^{~}_1(\mathbf{x}) + V^{~}_2 \left(\mathbf{x}^{~}_1, \mathbf{x}^{~}_2, \cdots , \mathbf{x}^{~}_n \right) + E^{~}_k \left(\mathbf{x},\mathbf{x}^{~}_1, \mathbf{x}^{~}_2,\cdots , \mathbf{x}^{~}_n \right)}{4\pi \left(\bm{\mathfrak{R}}^2_{\mathrm{particle}}\right)\nu'_{\mathrm{vib}}\nu^{~}_{\mathrm{rot}}}
						= \frac{E - E^{~}_{0}}{4\pi \left(\bm{\mathfrak{R}}^2_{\mathrm{particle}}\right)\nu'_{\mathrm{vib}}\nu^{~}_{\mathrm{rot}}}, 
				\end{align}
where $\mathbf{x} = \mathbf{R} + \mathbf{r}$ and $\mathbf{x}^{~}_n = \mathbf{R}^{~}_n + \mathbf{r}^{~}_n$. $V^{~}_1(\mathbf{x})$ is the Coulomb potential of the electron relatively to the nucleus, $V^{~}_2\left(\mathbf{x}^{~}_1, \mathbf{x}^{~}_2, \cdots , \mathbf{x}^{~}_n \right)$ is the Coulomb potential of an electron relative to the $n$ other electrons, and $E^{~}_k\left(\mathbf{x}, \mathbf{x}^{~}_1, \mathbf{x}^{~}_2,  \cdots , \mathbf{x}^{~}_n \right)$ is kinetic energy relative to both the nucleus and the $n$ other electrons.

Afterward if the particle undergoes emitting or absorbing processes with transition energy $\pm\Delta E_{\mathrm{transition}}$, the velocity field will change to keep the invariance of mass and Eq. (\ref{9}) becomes
				\begin{align*}
					m = \frac{E - E^{~}_0 \pm\Delta E_{\mathrm{transition}}}{4\pi \left(\bm{\mathfrak{R}}^2_{\mathrm{particle}}\right)\nu''_{\mathrm{vib}}\nu^{~}_{\mathrm{rot}}},
				\end{align*}
and the system will occupy a new state
				\begin{align}
					\label{10}
					m	= \frac{E - V^{~}_1(\mathbf{x}') + V^{~}_2 \left(\mathbf{x}'_1, \mathbf{x}'_2, \cdots , \mathbf{x}'_n \right) + E^{~}_k \left(\mathbf{x}',\mathbf{x}'_1, \mathbf{x}'_2,\cdots , \mathbf{x}'_n \right)}{4\pi \left(\bm{\mathfrak{R}}^2_{\mathrm{particle}}\right)\nu''_{\mathrm{vib}}\nu^{~}_{\mathrm{rot}}}
						= \frac{E - E_{0}'}{4\pi \left(\bm{\mathfrak{R}}^2_{\mathrm{particle}}\right)\nu''_{\mathrm{vib}}\nu^{~}_{\mathrm{rot}}},
				\end{align}											
$E^{~}_0$ and $E'_0$	 are the energies that have been transformed by the particle to create any physical system. When any transition process takes place with either emitting or absorbing processes, so the atom will rearrange itself through the change of potential terms and/or kinetic terms and the speed of random motion in the quantum-sized volume. Eqs. (\ref{9}) and (\ref{10}) represent the invariant mass principle when a transition process takes place, and propose that every $\Delta E^{~}_{\mathrm{transition}}$ can be generated by any combination of both a set $\left(V^{~}_1, V^{~}_2,\,\mathrm{and}\,E^{~}_k \right)$ and every combination of the subset of $\left(V^{~}_1, V^{~}_2,\,\mathrm{and}\,E^{~}_k \right)$, whereas every combination accompanying emitting or absorbing process will describe every possibility of interaction potential terms or coupling terms in the atom system. For the realistic model of an atom in which its concentrated-energy also represents the existence of environment expressed by fluctuating-field so that invariant mass principle requires that:
				\begin{align}
					\label{11}
					m 	&= \frac{E - E^{~}_0 + V^{~}_{\mathrm{noise}}}{4\pi \left(\bm{\mathfrak{R}}^2_{\mathrm{particle}}\right)\nu'_{\mathrm{vib}}\nu^{~}_{\mathrm{rot}}} \nonumber \\
						&= \frac{E - E^{~}_0 + V^{~}_{\mathrm{noise}} \pm \left(\Delta E \pm \delta E\right)_{\mathrm{transition}}}{4\pi \left(\bm{\mathfrak{R}}^2_{\mathrm{particle}}\right)\nu''_{\mathrm{vib}}\nu^{~}_{\mathrm{rot}}} \nonumber \\
						&= \frac{E - V^{~}_1(\mathbf{x}') + V_2 \left(\mathbf{x}'_1, \mathbf{x}'_2, \cdots , \mathbf{x}'_n \right) + E^{~}_k \left(\mathbf{x}',\mathbf{x}'_1, \mathbf{x}'_2,\cdots , \mathbf{x}'_n \right) + V'_{\mathrm{noise}}}{4\pi \left(\bm{\mathfrak{R}}^2_{\mathrm{particle}}\right)\nu''_{\mathrm{vib}}\nu^{~}_{\mathrm{rot}}} \nonumber \\
						&= \frac{E - E'_0 + V'_{\mathrm{noise}}}{4\pi \left(\bm{\mathfrak{R}}^2_{\mathrm{particle}}\right)\nu''_{\mathrm{vib}}\nu^{~}_{\mathrm{rot}}}.
				\end{align}
				\end{widetext}				 			
Thus \,$\left(\Delta E \pm \delta E\right)_{\mathrm{transition}}$ \,can\, be \, transformed \,into any combination of either a set $\left(V^{~}_1, V^{~}_2, E^{~}_k\right.$, and $\left.V_{\mathrm{noise}}\right)$ and every combination of the subset of $\left(V^{~}_1, V^{~}_2,E^{~}_k,\,\mathrm{and} \, V^{~}_{\mathrm{noise}}\right)$ \,\,where every combination accompanying the emitting or absorbing pro\-cess\-es will describe every possibility of the interaction potential terms or coupling terms, and connection between $V^{~}_{\mathrm{noise}}$ and $V'_{\mathrm{noise}}$ has been considered as either markovian or non-markovian processes.

				\item $1/\nu'_{\mathrm{vib}} = T'$ and $1/\nu''_{\mathrm{vib}} = T''$ are the period fields of the randomly vibrating-movement of the particle in the quantum-sized volume and it will play a role as the time interval unit for previous and new states.
				
				On of the most important things from invariant mass principle is that it is possible for every physical system to evolve in time without undergoing an emission or absorption process as long as the changes do not alter the total $E_0$ of each particle of the system, or the energy particle is constant, or even can take place by emission or absorption processes but the total amount of $E_0$ does change on average.
						
			\end{enumerate}

From both proposal of the fundamental motion of the particle and the new concept of mass, it seems that there are two possible mechanisms that act as the sources of unpredictability of measurement outcome:
			\begin{enumerate}
				\item Random motion taking place in the quantum-sized volume. (Internal random motion makes physical observables asserted and coupled with position will have statistical features).
				
				\item Macroscopic fluctuations, such as all process that happen in the universe including instrument states. Although the average macroscopic fluctuation do not change the stationary state of particles (any physical system), they will contribute to the outcome probabilities.
			\end{enumerate}

		\section{\label{sec:4}THE DIFFUSION CONSTANT IN THE TRANSITION PROCESS}

To derive the diffusion constant of the particle, we consider a transition process where the particle emits energy to coincide with doing a diffusion process taking place in the quantum-sized volume. First, we consider that the particle is in a state without any kind of interaction or any kind of system or environment perspective, so that the relation between the concentrated-energy and the value of mass will be
			\begin{align*}
				m	= \frac{E}{4 \pi \left( \bm{\mathfrak{R}}^2_{\mathrm{particle}} \right) \nu^{~}_{\mathrm{vib}} \nu^{~}_{\mathrm{rot}}}
					= \frac{E}{\mathbf{c}^2}.
			\end{align*}			 	
When the particle makes the transition process to occupy a new state to form a new simple physical system consisting of two particles with no kinetic term of the quantum-sized volume, so we can write the process in a simple equation as
			\begin{align}
				\label{12}
				m	= \frac{E - \Delta E^{~}_{\mathrm{transition}}}{4 \pi \left( \bm{\mathfrak{R}}^2_{\mathrm{particle}} \right) \nu'_{\mathrm{vib}} \nu^{~}_{\mathrm{rot}}}
					= \frac{E - V(\mathbf{R} + \mathbf{r}) + E^{~}_{k}(\mathbf{r})}{4 \pi \left( \bm{\mathfrak{R}}^2_{\mathrm{particle}} \right) \nu'_{\mathrm{vib}} \nu^{~}_{\mathrm{rot}}}.
			\end{align}
$-V(\mathbf{R} + \mathbf{r})$ shows the two particles have different types of charge and $E_k(\mathbf{r})$ is the kinetic energy of the particle in the quantum-sized volume. Because the particle undergoes a diffusion process in the quantum-sized volume, work that is equivalent to the displacement by diffusion process is expressed by
			\begin{align}
				\label{13}
				m \left(\frac{\mathrm{d}^2\mathbf{r}}{\mathrm{d}t^2} \right) \cdot \mathbf{r}
					&= -m\left[ 4 \pi \left( \bm{\mathfrak{R}}^2_{\mathrm{particle}} \right) \nu'_{\mathrm{vib}} \nu^{~}_{\mathrm{rot}} \right]  \nonumber \\
					&= \Delta E^{~}_{\mathrm{transition}} - E.
			\end{align}											
Negative work means that the emission process is generated by the environment, such as the friction process in the perspective of Langevin's equation. The process in Eq. (\ref{13}) can be considered from Langevin's equation with force
			\begin{subequations}			
			\begin{align}
				\label{14a}
				\left( \frac{\mathrm{d}^2 \mathbf{r}}{\mathrm{d}t^2} \right)
					&= -\frac{\xi}{m}\left(\frac{\mathrm{d}\mathbf{r}}{\mathrm{d}t}\right) + \frac{1}{m}\mathbf{f}_{\mathrm{random}}, \\
				\label{14b}
				m\left( \frac{\mathrm{d}^2 \mathbf{r}}{\mathrm{d}t^2} \right)
					&= -\xi\left(\frac{\mathrm{d}\mathbf{r}}{\mathrm{d}t}\right) + \mathbf{f}_{\mathrm{random}}.
			\end{align}
Using the definition of work
			\begin{align}
				\label{14c}
				m\left( \frac{\mathrm{d}^2 \mathbf{r}}{\mathrm{d}t^2} \right) \cdot \mathbf{r}
					&= -\xi\left(\frac{\mathrm{d}\mathbf{r}}{\mathrm{d}t}\right) \cdot \mathbf{r} + \mathbf{f}_{\mathrm{random}} \cdot \mathbf{r},
			\end{align}
with
			\begin{align}
				\label{14d}
				\frac{1}{2}m \left(\frac{\mathrm{d}^2\mathbf{r}^2}{\mathrm{d}t^2}\right)
					= m\left( \frac{\mathrm{d}^2\mathbf{r}}{\mathrm{d}t^2} \right) \cdot \mathbf{r}
						+ m \left(\frac{\mathrm{d}\mathbf{r}}{\mathrm{d}t}\right)^2,
			\end{align}
we obtain
			\begin{align}
				\label{14e}
				m\left( \frac{\mathrm{d}^2\mathbf{r}}{\mathrm{d}t^2} \right) \cdot \mathbf{r}
					= -\frac{\xi}{2}\left(\frac{\mathrm{d}\mathbf{r}^2}{\mathrm{d}t}\right)
					= \frac{m}{2}\left(\frac{\mathrm{d}^2\mathbf{r}^2}{\mathrm{d}t^2}\right) 
						- m\left(\frac{\mathrm{d\mathbf{r}}}{\mathrm{d}t}\right)^2.
			\end{align}
			\end{subequations}
In Eq. (\ref{14a}), $\mathbf{f}$ is the random forces that are produced by the collision between the particle and a medium, but because random motion, in our model, is an intrinsic properties of the particle, $\mathbf{f} = 0$. This result is quantitatively no different from classical viewpoint because the length on average is null $\left( \left\langle \mathbf{r} \right\rangle = 0 \,\mathrm{or}\,\left\langle \mathbf{f} \cdot \mathbf{r} \right\rangle = 0 \right)$, while $\xi$ is a friction factor depending on geometry of molecules. We express $E$ in Eq. (\ref{13}) within differential form
			\begin{align}
				\label{15}
				E 	&= \left\langle m \, 4\pi \bm{\mathfrak{R}}^2_{\mathrm{particle}}\nu^{~}_{\mathrm{vib}}\nu^{~}_{\mathrm{rot}}\right\rangle \nonumber \\
					&= \left\langle m \mathbf{c}^2 \right\rangle \nonumber \\
					&= m \mathbf{c}^2 \nonumber \\
					&= m \left(\frac{\mathrm{d}\mathbf{r}}{\mathrm{d}t}\right)^2. 
			\end{align}
Because the geometry factor $\xi$ is associated with $\nu'_{\mathrm{vib}}$ determining the amount of either emission or absorption energies, furthermore, we represent the vector field $\bm{\mathfrak{R}}$ with a displacement vector field $\mathbf{r}$ so that we can express
			\begin{align}
				\label{16}
				-m &\left(4\pi \bm{\mathfrak{R}}^2_{\mathrm{particle}}\,\nu'_{\mathrm{vib}}\nu^{~}_{\mathrm{rot}}\right)\nonumber \\
					&\qquad= -m\left(4\pi \bm{\mathfrak{R}}^{~}_{\mathrm{particle}}\nu'_{\mathrm{vib}}\frac{\bm{\mathfrak{R}}^{~}_{\mathrm{particle}}}{T_{\mathrm{rot}}}\right) \nonumber\\
					&\qquad\approx -m\left(4 \pi \mathbf{r} \, \nu'_{\mathrm{vib}} \frac{\mathrm{d}\mathbf{r}}{\mathrm{d}t}\right) \nonumber \\
					&\qquad\approx -2\pi\, m \,\nu'_{\mathrm{vib}}\frac{\mathrm{d}\mathbf{r}^2}{\mathrm{d}t}.
			\end{align}
where $T^{~}_{\mathrm{rot}}$ is the period of internal angular motion.

Therefore we may write Eq. (\ref{13}) as
			\begin{align}
				\label{17}
				\Delta E^{~}_{\mathrm{transition}} - m\left(\frac{\mathrm{d}\mathbf{r}}{\mathrm{d}t}\right)^2
					= -2\pi m\nu'_{\mathrm{vib}} \frac{\mathrm{d}\mathbf{r}^2}{\mathrm{d}t}.
			\end{align}					
Eqs. (\ref{14e}) and (\ref{17}) represent the same process, therefore $\Delta E_{\mathrm{transition}}$ will be 
			\begin{align}
				\label{18}
				\Delta E_{\mathrm{transition}}
					= \frac{m}{2}\left(\frac{\mathrm{d}^2\mathbf{r}^2}{\mathrm{d}t^2}\right),
			\end{align}	
and
			\begin{align}
				\label{19}
				\xi = 4m\pi \nu'_{\mathrm{vib}}.
			\end{align}	
Using energy relation, we have
			\begin{align}
				\label{20}
				E 	= \left\langle m\mathbf{c}^2\right\rangle = m \mathbf{c}^2
					= \left\langle m \left( \frac{\mathrm{d}\mathbf{r}}{\mathrm{d}t} \right)^2\right \rangle 
					= h\nu.
			\end{align}									

From the invariant mass principle, it is clear that the emitting and absorbing processes will coincide with the decrease or increase of the vibration-frequency field, therefore the emission and absorption processes will correspond with positive or negative values of the change of $\nu'_{\mathrm{vib}}$ or $\xi$ (excitation and de-excitation processes can be represented by the decrease or the increase of the speed of internal random motion). Use a new notation $\alpha = \left\langle \mathrm{d}\mathbf{r}^2 \right\rangle / \mathrm{d}t$, and applying the condition that the random vibrating-frequency of the charged particle and the frequency of its field must be the same, so Eq. (\ref{14e}) and (\ref{17}) can be written as
			\begin{align}
				\label{21}
				\left(\frac{\mathrm{d}\alpha}{\mathrm{d}t}\right) + \frac{\xi}{m} \alpha
					= \frac{2h \nu}{m}.
			\end{align}
Eq. (\ref{21}) has a general solution
			\begin{align}
				\label{22}
				\alpha = \frac{2h \nu}{\xi} + C \exp\left(-\frac{\xi}{m}t\right).
			\end{align}
Because the time interval of the transition process will much higher than the period of the vibration process taken place in the quantum-sized volume or $t \gg \xi/m$ or $\exp\left(-\xi t/m\right) \rightarrow 0$, so we obtain
			\begin{align}
				\label{23}
				\left\langle \mathbf{r}^2 \right\rangle
					= \frac{2h\nu}{\xi}t.
			\end{align}
Whereas according to Stokes-Einstein-Sutherland equation, diffusion constant is
			\begin{align}
				\label{24}
				\beta = \lim\limits_{t \rightarrow \infty} \frac{1}{2 \Delta t} \left \langle \mathbf{r}^2 \right\rangle.
			\end{align}
Using the assumption that the time interval of transition process will much higher than the period of the vibration process or taking the perspective that a process taking place is sufficiently long time so that Eq. (\ref{24}) is
			\begin{align}
				\label{25}
				2 t \beta 
					&= \left \langle\mathbf{r}^2 \right\rangle \nonumber \\
				\beta
					&= \frac{\left\langle \mathbf{r}^2 \right\rangle}{2t}.
			\end{align}
Using Eq. (\ref{23}) and because $\xi = 4 \pi m \nu$, we find
			\begin{align}
				\label{26}
				\beta 
					&= \frac{h}{4\pi m} \nonumber \\
					&= \frac{\hslash}{2m}.
			\end{align}

	\section{\label{sec:5}DERIVING SCHR\"{O}DINGER EQUATION}	
		
We \,consider \,the \,dynamics \,aspect \,of \,our model and make interpretations due to the meaning behind Schr\"{o}dinger equation. Applying Eq. (\ref{1}) and (\ref{2}) in\-to Eq. (\ref{7}) we have
		\begin{align}
			\label{27}
			\frac{1}{m}\mathbf{F}_{\mathrm{external}} - \frac{1}{m}\nabla V
				&= \frac{\partial^2 \mathbf{R}}{\partial t^2} + \frac{1}{2}\frac{\partial}{\partial t} \left(\mathbf{b} + \mathbf{b}^{~}_*\right) \nonumber \\
				&\qquad + \frac{1}{2}\left[\left(\mathbf{b} + \mathbf{b}^{~}_*\right) \cdot \nabla\right] \frac{\partial \mathbf{R}}{\partial t} \nonumber \\
				&\qquad + \frac{1}{2}\left(\mathbf{b} \cdot \nabla \right)\mathbf{b}^{~}_* + \frac{1}{2}\left(\mathbf{b}^{~}_* \cdot \nabla\right)\mathbf{b} \nonumber \\
				&\qquad	- \frac{\hslash}{4m}\nabla^2 \left(\mathbf{b} - \mathbf{b}^{~}_*\right).
		\end{align}				
We rewrite Eq. (\ref{27}) as
		\begin{align}
			\label{28}
			\frac{1}{m}\mathbf{F}^{~}_{\mathrm{external}} - \frac{1}{m} \nabla V
				&= \frac{\partial \mathbf{v}}{\partial t} + \frac{1}{2} \frac{\partial }{\partial t} \left(\mathbf{b} + \mathbf{b}^{~}_*\right) \nonumber \\
				&\qquad + \frac{1}{2}\left[\left(\mathbf{b} + \mathbf{b}^{~}_*\right)\cdot \nabla\right]\mathbf{v} \nonumber \\ 
				&\qquad + \frac{1}{2}\left(\mathbf{b}\cdot \nabla\right)\mathbf{b}^{~}_* \nonumber \\
				&\qquad + \frac{1}{2}\left(\mathbf{b}^{~}_* \cdot \nabla\right) \mathbf{b} \nonumber \\
				&\qquad - \frac{\hslash}{4m}\nabla^2 \left(\mathbf{b} - \mathbf{b}^{~}_*\right) ,
		\end{align}				
where $\partial\mathbf{R}/\partial t = \mathbf{v}$. This result is different from the Nelson model which does not include $\partial \mathbf{x}(t)/\partial t = \mathbf{v}$, in the proposal model, $\partial \mathbf{x}(t)/\partial t = \partial \mathbf{R}(t)/\partial t = \mathbf{v}$ represents the motion of a quantum-sized volume, however $\partial \mathbf{x}(t)/\partial t = \mathbf{v}$ may also describe the motion of vacuum (medium).

According to our concept of mass, the particle must undergo random motion in the quantum-sized volume, thus the probability density to find a particle at any point in the quantum-sized volume will directly depend on the speed of the random motion and distribution of the random-velocity field for stationer state as well as depend on the diffusion process taking place in the quantum-sized volume for transition process. Because the proposed definition of mass covers the whole space of the quantum-sized volume, the distribution of mass in the quantum-sized volume is interchangeable with the probability distribution to find a particle in the quantum-sized volume, and how the probability density changes from one state to another can be interchangeably viewed as the change of either the probability density or the mass density or the charge density. Although the density of the mass and the charge of particle is always $m$ and $q$, which are equivalent to the condition that the total probability to find a particle in of the entire space of the quantum-sized volume is one.

Defining $\rho\left(\mathbf{r}', t\right)$ as the probability density to find the particle in the quantum-sized volume, so that $\rho\left(\mathbf{r}',t\right)$ obeys
		\begin{align}
			\label{29}
			\int \rho\left( \mathbf{r}', t \right)\, \mathrm{d}^3 r'  = 1.
		\end{align}		
This describes the probability of finding the particle in the entire space of the quantum-sized volume, which must be one. If we define $\rho^{~}_e\left( \mathbf{r}' , t \right)$ and $\rho_m\left( \mathbf{r}' , t \right)$	as charge and mass densities in the quantum-sized volume respectively, so that they obey
		\begin{align}
			\label{30}
			\int \rho^{~}_e \left( \mathbf{r}' , t \right)\, \mathrm{d}^3 r' = q,
		\end{align}
		\begin{align}
			\label{31}
			\int \rho^{~}_m \left( \mathbf{r}' , t \right)\, \mathrm{d}^3 r' = m.
		\end{align}
From Eqs. (\ref{29}) and (\ref{30}) as well as (\ref{31}), we can define the connection among the probability, charge and mass densities by
		\begin{align}
			\label{32}
			\rho^{~}_e\left( \mathbf{r}' , t \right) = q  \rho\left( \mathbf{r}' , t \right),
		\end{align}
		\begin{align}
			\label{33}
			\rho^{~}_m\left( \mathbf{r}' , t \right) = m \rho\left( \mathbf{r}' , t \right).
		\end{align}

From hydrodynamics viewpoint, we can imagine the mass density or charge density or probability density as fluid density. As shown in session (3), the fluid analogy is presented by the Langevin equation containing the coefficient friction describing interaction between the Brownian particle and the fluid (medium) particle. Based on this viewpoint, we can associate the properties of the field of internal motion velocity and emission or absorption processes with mechanical properties and thermodynamical-statistical properties of fluid. For instance, we can associate internal mass density (spreading over the quantum-sized volume) with fluid density.

Internal random motion taking place in the quantum-sized volume, where the total mass and charge for the entire quantum-sized volume is always $m$ and $q$, makes us can consider a point in the quantum-sized volume as a particle of both mass $\rho^{~}_m\left(\mathbf{r}', t\right)$ and charge $\rho^{~}_e\left( \mathbf{r}', t \right)$. Because a particle of both mass $\rho^{~}_e\left( \mathbf{r}', t \right)$ and charge $\rho^{~}_e \left( \mathbf{r}', t \right)$ undergoes random motion where the density of the field of the internal random velocity is always relatively much higher than the time interval of measurement (observer), we will always see the current of the point particle with mass $\rho^{~}_m \left(\mathbf{r}', t\right)$ and charge $\rho^{~}_e\left( \mathbf{r}' , t \right)$ for every time interval as the current density that will obey the continuity equation for emission and absorption processes for mass and charge and probability densities
		\begin{subequations}
			\begin{align}
				\label{34a}
				\frac{\partial \rho^{~}_{m,e,p} \left( \mathbf{r}' , t \right)}{\partial t}
					&= -\nabla \cdot 
						\left[
							\left( \mathbf{b} + \mathbf{v} \right) \rho_{m, e, p} \left(\mathbf{r}',t \right)
						\right] \nonumber \\
					&\qquad + \frac{\hslash}{2m} \nabla^2 \rho^{~}_{m, e, p}\left( \mathbf{r}' , t \right), \\
				\label{34b}
				\frac{\partial \rho^{~}_{m,e,p} \left( \mathbf{r}' , t \right)}{\partial t}
					&= -\nabla \cdot 
						\left[
							\left( \mathbf{b}^{~}_* + \mathbf{v} \right) \rho_{m, e, p} \left(\mathbf{r}',t \right) 
						\right] \nonumber \\
					&\qquad	- \frac{\hslash}{2m} \nabla^2 \rho^{~}_{m, e, p}\left( \mathbf{r}' , t \right),
			\end{align}
		\end{subequations}
where index $m, e, p$ respectively refer to mass, charge, and probability. Eq. (\ref{34a}) and (\ref{34b}) are forward and backward modified-Fokker equations. Eqs. (\ref{34a}) and (\ref{34b}) yield
		\begin{align}
			\label{35}
			\frac{\partial \rho}{\partial t} = -\nabla \cdot \left[\left( \bm{\upsilon} + \mathbf{v} \right)\rho\right],
		\end{align}
and
		\begin{align}
			\label{36}
			\mathbf{u} = \beta \frac{\nabla \rho}{\rho},
		\end{align}
where $\bm{\upsilon}$ is
		\begin{align}
			\label{37}
			\bm{\upsilon} = \frac{1}{2} \left( \mathbf{b} + \mathbf{b}^{~}_* \right),
		\end{align}
and we call $\bm{\upsilon}$ as the transition velocity. While $\mathbf{u}$ is defined by
		\begin{align}
			\label{38}
			\mathbf{u} = \frac{1}{2} \left( \mathbf{b} - \mathbf{b}^{~}_* \right),
		\end{align}
and we cal $\mathbf{u}$ as the stationer velocity. Computing $\partial \mathbf{u}/\partial t$ and applying (\ref{35}), we obtain
		\begin{align}
			\label{39}
			\frac{\partial \mathbf{u}}{\partial t}
				= -\frac{\hslash}{2m} \nabla \left[\nabla \cdot \left( \bm{\upsilon} + \mathbf{v}\right) \right]
					-\nabla\left[\mathbf{u} \cdot \left(\bm{\upsilon} + \mathbf{v}\right)\right].
		\end{align}
Applying (\ref{37}) and (\ref{38}) to (\ref{28}), we find
		\begin{align}
			\label{40}
			\frac{\partial \left(\bm{\upsilon} + \mathbf{v}\right)}{\partial t}
				&= -\frac{1}{m}\left( \nabla V - \mathbf{F}^{~}_{\mathrm{external}}\right) 
					- \left( \bm{\upsilon} \cdot \nabla \right) \left( \bm{\upsilon} + \mathbf{v} \right) \nonumber \\
				&\qquad + \left( \mathbf{u} \cdot \nabla \right)\mathbf{u}
					+ \frac{\hslash}{2m}\nabla^2 \mathbf{u}.
		\end{align}

		\subsection{\label{sec:5-1}The real time-independent Schr\"{o}dinger equation}
	
We consider the stationer state (i.e. no transition process). From the description of the stationer state in Eq. (\ref{8}), the stationer state requires
			\begin{align*}
				\mathbf{F}_{\mathrm{external}} = 0,
			\end{align*}
whereas from definition of the stationer state at Eqs. (\ref{1}) and (\ref{2}), a stationery state requires
			\begin{align}
				\label{41}
				\mathbf{b} = -\mathbf{b}_*,
			\end{align}
or $\bm{\upsilon} = 0$ so that $\partial\bm{\upsilon}/\partial t = 0$, furthermore $\partial \mathbf{u} / \partial t = 0$ so that $\rho$ and $\mathbf{u}$ are independent of $t$. We have two cases for stationer states that are $\mathbf{v} = 0$ and $\mathbf{v} \neq 0$. For $\mathbf{v} = 0$ case, by applying the stationer conditions to Eq. (\ref{39}) and (\ref{40}), we obtain
			\begin{align}
				\label{42}
				\left( \mathbf{u} \cdot \nabla \right)\mathbf{u} + \frac{\hslash}{2m}\nabla^2 \mathbf{u}
					= \frac{1}{m}\nabla V \\
				\label{43}
				\frac{1}{2}\nabla\mathbf{u}^2 + \frac{\hslash}{2m} \nabla \left( \nabla \cdot \mathbf{u} \right)
					= \frac{1}{m}\nabla V.
			\end{align}						

According to the invariant mass principle, the total concentrated-energy of the particle before the physical system exists is $E = m \mathbf{c}^2$, and the system is the representation of how much the concentrated-energy has transformed to present potential terms, kinetic terms and the decrease of random motion to less than the speed of light. Furthermore, the transition energy represents how the system changes from any combination of potential and/or kinetic and random velocity terms to another combination. Therefore, the total amount of kinetic and potential energies at every stationer state must be equal to the total energy that is transformed by the particle, called $E^{~}_0$. If we integrate Eq. (\ref{43}), the constant of integration should be the negative of $E^{~}_0$. We have
			\begin{align}
				\label{44}
				\frac{1}{m}E^{~}_0 + \frac{\hslash}{2m}\left( \nabla \cdot \mathbf{u} \right)
					= -\frac{1}{2}\mathbf{u}^2 + \frac{1}{m}V. 
			\end{align}

We notice $\mathbf{u} = \mathbf{b} = -\mathbf{b}_*$ corresponds with the emission and absorption processes, and that the $\pm\left(\hslash/2m\right)\left( \nabla \cdot \mathbf{u} \right)$ terms represent either emitted-energy or absorbed-energy when the particle undergoes transition process. Therefore, $\left(\hslash/2\right)\left( \nabla \cdot \mathbf{u} \right)$ must be represent the difference between energy level of stationery state. Since $E$ is constant and $\left(\hslash/2\right)\left( \nabla \cdot \mathbf{u} \right)$ has definite value, every state must be quantified by both space variables due to the potential and intrinsic variable that represents random motion in the quantum-sized volume. Following Nelson's work, Eq. (\ref{44}) will be time independent Schr\"{o}dinger equation
			\begin{align}
				\label{45}
				\left[-\left(\frac{\hslash^2}{2m}\right) \nabla^2 + V\right] \Psi
					= E^{~}_0 \Psi
			\end{align}
with
			\begin{align}
				\label{46}
				\Psi = e^R,
			\end{align}
where $R$ in Eq. (\ref{46}) satisfy $R = \left(\ln \rho\right)/2$.

For $\mathbf{v} \neq 0$, form Eqs. (\ref{39}) and (\ref{40}), we obtain
			\begin{align}
				\label{47}
				\beta \nabla^2 \mathbf{v}
					= - \nabla \left( \mathbf{u} \cdot \mathbf{v} \right),
			\end{align}
			\begin{align}
				\label{48}
				\mathbf{a} = -\left( \mathbf{u} \cdot \nabla \right)\mathbf{u} - \beta \nabla^2 \mathbf{u}.
			\end{align}
Eqs. (\ref{47}) and (\ref{48}) determine that the energy of the emission or absorption processes is contributed to by the change of both the intrinsic and the quantum-sized volume velocities and can be presented by single equation
			\begin{align}
				\label{49}
				\left[-\left(\frac{\hslash^2}{2m}\right) \nabla^2 + V\right] \Psi
					= E^{~}_v \Psi,
			\end{align}		 
where $\Psi = e^{R + iS}$, $\nabla S = m\mathbf{v}/\hslash$, $E^{~}_v = E^{~}_0 + E^{~}_k$, and $E^{~}_k = mv^2/2$. Eqs. (\ref{47}) and (\ref{48}) are the imaginary and real parts of Eq. (\ref{49}) depicting that the emergence of the kinetic term will be compensated by the decrease of the amount of concentrated-particle energy.

		\subsection{\label{sec:5-2}The real time-dependent Schr\"{o}dinger equation}

There are two kinds of general time-dependent cases that are physical systems without or with motion of the quantum-sized volume. For systems with $\mathbf{v} = 0$, Eqs. (\ref{39}) and (\ref{40}) become
			\begin{align}
				\label{50}
				\frac{\partial \mathbf{u}}{\partial t}
					= -\beta \nabla \cdot \bm{\upsilon} - \nabla \left( \mathbf{u} \cdot \bm{\upsilon} \right),
			\end{align}
			\begin{align}
				\label{51}
				\frac{\partial \bm{\upsilon}}{\partial t}
					= \mathbf{a} - \left( \bm{\upsilon} \cdot \nabla \right)\bm{\upsilon}
								 + \left( \mathbf{u} \cdot \nabla \right)\mathbf{u}
								 + \frac{\hslash}{2m} \nabla^2 \mathbf{u}.
			\end{align}
It has been shown by \cite{EN-1966} that Eq. (\ref{50}) and (\ref{51}) are equivalent to the Schr\"{o}dinger equation:
			\begin{align}
				\label{52}
				i\hslash \frac{\partial \Psi}{\partial t}
					= -\frac{\hslash^2}{2m}\nabla^2 \Psi + V_{\mathrm{net}}\Psi,
			\end{align}
where $V^{~}_{\mathrm{net}} = V + V^{~}_{\mathrm{external}}$, and $\Psi = e^{R + iS}$ where $\nabla S = m\bm{\upsilon}/\hslash$.

We can see that $\bm{\upsilon}$ in Eqs. (\ref{50}) and (\ref{51}) also presented in the modified-Fokker-Planck Eqs. (\ref{34a}) and (\ref{34b}) thus indicate that it is a field. In our proposal, the probability density to find a particle in the quantum-sized volume is equivalent to the mass density of the particle in the quantum-sized volume where it will be determined by the amount of concentrated-energy. While the kinetic term is one of the features of the physical system which represents the change in the amount of concentrated energy, hence the velocity of the quantum-sized volume will be a physical quantity which governs the probability of finding a particle in the quantum-sized volume. This means that when the quantum-sized volume moves from one state with certain velocity to another state with different velocity accompanied by the change of velocity of internal random motion, so the probability density will change simultaneously as well. The physical mechanism can be also understood from another perspective in that the quantum-sized volume will move from one point to another point with a certain probability. From a stochastic mechanics viewpoint, this can be considered as the Brownian motion of the quantum-sized volume from one point to another point that generates the field property of $\bm{\upsilon}$.

For a system with $\mathbf{v} \neq 0$, Eqs. (\ref{39}) and (\ref{40}) will be equivalent to a new form of the Schr\"{o}dinger equation
			\begin{align}
				\label{53}
				i\hslash \frac{\partial \Psi}{\partial t}
					&= -\frac{\hslash^2}{2m}\nabla^2\Psi \nonumber \\
					&\qquad + \left[ V^{~}_{\mathrm{net}} + E^{~}_k  
								+ \int m \left( \bm{\upsilon} \nabla \cdot \mathbf{v} \right)\,\mathrm{d}^3 r' \right]\Psi,
			\end{align}
where $\Psi = e^{R + iS}$, with $\nabla S = m\left( \bm{\upsilon} + \mathbf{v}\right)/\hslash$, whereas $E^{~}_k = mv^2/2$ and $\mathrm{d}^3 r'$ is the volume element of the quantum-sized volume. For a special case when $\mathbf{u}$ is a solenoidal vector field ($\nabla \cdot \mathbf{v} = 0$) then the Eq. (\ref{53}) becomes
			\begin{align}
				\label{55}
				i\hslash\frac{\partial \Psi}{\partial t}
					= -\frac{\hslash^2}{2m}\nabla^2\Psi + \left( V_{\mathrm{net}} + E_k\right)\Psi.
			\end{align}
In this case, the particle is not in any physical system (the particle is not governed by any potential).

The Schr\"{o}dinger equations for the cases $\mathbf{v} = 0$ and $\mathbf{v} \neq 0$ are respectively
			\begin{align}
				\label{56}
				i\hslash\frac{\partial \Psi}{\partial t}
					= -\frac{\hslash^2}{2m}\nabla^2\Psi + V_{\mathrm{external}}\Psi,
			\end{align}
			\begin{align}
				\label{57}
				i\hslash \frac{\partial \Psi}{\partial t}
					&= -\frac{\hslash^2}{2m}\nabla^2\Psi \nonumber \\
					&\qquad + \left[ V^{~}_{\mathrm{net}} + E^{~}_k  
								+ \int m \left( \bm{\upsilon} \nabla \cdot \mathbf{v} \right)\,\mathrm{d}^3 r' \right]\Psi,
			\end{align}
or for a special case when $\mathbf{u}$ is a solenoidal vector field ($\nabla \cdot \mathbf{v} = 0$), Eq. (\ref{53}) becomes		
			\begin{align}
				\label{58}
				i\hslash\frac{\partial \Psi}{\partial t}
					= -\frac{\hslash^2}{2m}\nabla^2\Psi + \left( V_{\mathrm{external}} + E_k\right)\Psi.
			\end{align}
			
		\subsection{\label{sec:5-3}The origin of spin on the new model of stochastic interpretation}	

Recalling the modified-Fokker-Planck equations
			\begin{align*}
				\frac{\partial \rho}{\partial t}
					&= -\nabla \cdot \left[\left(\mathbf{b} + \mathbf{v}\right)\rho\right] 
						+ \frac{\hslash}{2m}\nabla^2\rho,		\\
				\frac{\partial \rho}{\partial t}
					&= -\nabla \cdot \left[\left(\mathbf{b}_* + \mathbf{v}\right)\rho\right] 
						- \frac{\hslash}{2m}\nabla^2\rho.
			\end{align*}
Spin, in our proposal, is connected with the internal random motion that takes place in the quantum-sized volume. The particle moves by random walking with certain velocities will change its velocity field when every transition taking place coincides with forward or backward diffusions for either emitting or absorbing processes. The change of the velocity field of the random motion in the quantum-sized volume (limited motion) will generate internal spinning motion. Each transition process (emission and absorption processes), in our proposal, is represented in Eqs. (\ref{34a}) and (\ref{34b}). Eq. (\ref{34a}) describes if the particle occupies a certain general stationer state with both certain random velocity field $\mathbf{b}$ and the certain quantum-sized volume velocity $\mathbf{v}$, so when the transition process coincides with the diffusion process (emission process), the particle will evolve to a new stationer state with the velocity of random motion $\mathbf{b}_*$.  While Eq. (\ref{34b}) describes backward process corresponding to the absorption processes represented by negative coefficient diffusion. Because the change of the internal velocity field takes place in the quantum-size volume, it is very clear that this mechanism will generate spinning-motion perspective. The velocity field representing the change of internal random motion is shown by Eq. (\ref{36})
			\begin{align*}
				\mathbf{u} = \frac{\hslash}{2m}\frac{\nabla \rho}{\rho}.
			\end{align*}					

Now, we define the normal-direction field to $\mathbf{u}$ (the unit vectors that is perpendicular to the plane of $\mathbf{b}$ and $\mathbf{b}_*$) as the direction field of internal spinning-motion and we can call it $\hat{\mathbf{s}}$. Using a description of $\hat{\mathbf{s}}$, we can approximately define velocity fields $\mathbf{b}$ and $\mathbf{b}_*$ with $\mathbf{u}$ and $\hat{\mathbf{s}}$, that are
			\begin{align}
				\label{59}
				\mathbf{b}
					\approx \frac{\hslash}{2m}\frac{\nabla\rho}{\rho} \times \hat{\mathbf{s}},
			\end{align}
			\begin{align}
				\label{60}
				\mathbf{b}_*
					\approx \frac{\hslash}{2m}\frac{\nabla \rho}{\rho} \times \left(-\hat{\mathbf{s}}\right).	
			\end{align}
For the stationer state, we rewrite the current density in the Eqs. (\ref{34a}) and (\ref{34b}) with
			\begin{align}
				\label{61}
				\mathbf{J}
					= \left[ \left(\frac{\nabla \rho}{m \rho} \times \hat{\mathbf{s}}\right) + \mathbf{v}\right] \rho,
			\end{align}
			\begin{align}
				\label{62}
				\mathbf{J}
					= \left[-\left(\frac{\nabla \rho}{m \rho} \times \hat{\mathbf{s}}\right) + \mathbf{v}\right] \rho,
			\end{align}
where $\mathbf{s} = \left(\hslash/2\right)\hat{\mathbf{s}}$ and $\mathbf{J}$ is the current density. Since $\hat{\mathbf{s}}$ is perpendicular to $\mathbf{u} = \left(\hslash/2m\right)\left(\nabla \rho\right)/\rho$  so that
			\begin{align}
				\label{63}
				\left[\frac{\nabla \rho}{m \rho} \times \mathbf{s} \right]^2
					= \left(\frac{\nabla \rho}{m \rho}\right)^2 \mathbf{s}^2.
			\end{align}
It is known from Clifford algebra to Dirac theory \cite{ER-GS-1998}.

		\subsection{\label{sec:5-4}The relation among time interval, magnetic and electric fields as well as mass}

In this section, we will introduce a new approach to understand the origin of charge (fields) via the existence of magnetic and electric fields based on the proposed picture of mass. Mass, as having been described in the previous section, when the interaction does not exist, will connect to the amount of concentrated-energy contained by particle following equation
			\begin{align*}
				m 	= \frac{E}{4\pi\left( \bm{\mathfrak{R}}^2_{\mathrm{particle}} \right) 
									\nu^{~}_{\mathrm{vib}} \nu^{~}_{\mathrm{rot}}}
					= \frac{E}{\mathbf{c}^2}.
			\end{align*}
When an interaction exist to create any physical system -- for instance, to create the simple physical system that is only governed by the electric potential and kinetic energy -- the particle will emit energy through an emission process and thus cause the particle to evolve to occupy a state. In this process, the particle's mass will be always constant to obey the invariant mass principle
			\begin{align*}
				m 	= \frac{E - \Delta E^{~}_{\mathrm{transition}}}{4\pi\left( \bm{\mathfrak{R}}^2_{\mathrm{particle}} \right) 
																	\nu'_{\mathrm{vib}} \nu^{~}_{\mathrm{rot}}}
					= \frac{E - V(\mathbf{R} + \mathbf{r}) 
							+ E^{~}_k(\mathbf{r})}{4\pi\left( \bm{\mathfrak{R}}^2_{\mathrm{particle}} \right)
																	\nu'_{\mathrm{vib}} \nu^{~}_{\mathrm{rot}}}.	
			\end{align*}			 
Through this simple process, we can view the electric potential or electric field as a representation of the particle's tendency to have a minimum concentrated-energy after the emission process where to maintain a minimum energy, the two particles come closer to each other (for system that consist of two particles with opposite charge). Contrary to a system with two opposite charges, for simple system consisting of two particles with the same charge, the two particles will stay away from each other in order to maintain a minimum concentrated-energy. Thus, two particles with opposite charge, if interaction makes the perspective of kinetic energy between both references change while the perception on distance between them does not change, the perception on the change of kinetic energy perception (for instance, according to one frame of reference, the kinetic energy of another particle increase) will be responded as the loss of the particle's tendency to come closer to each other (the loss of electric potential), or it represents the emergence of magnetic potential (field) between them. We see the change of the perspective of kinetic energy between two references represents the transformation of fields. In the same way, for a case, a particle experience the change of perception on both distance and velocity in the same time (for example when the particle makes a transition process from one state to another state) then this perception will be responded as the emergence of electromagnetic field between two terms of reference. Thus, electric and magnetic fields are relatively phenomena between two terms of reference. To show this view, we consider a Hydrogen atom in a simple closed-system (we ignore the fluctuating-environment field). According to the invariant mass principle, an electron at any state will describe how much energy of the particle has been transformed to potentials, kinetic energy, and the resulting decrease of internal random velocity to a speed less than speed of light
			\begin{align}
				\label{64}
				m	&= \frac{E - E^{~}_0}{4\pi\left( \bm{\mathfrak{R}}^2_{\mathrm{particle}} \right)
										\nu'_{\mathrm{vib}} \nu^{~}_{\mathrm{rot}}} \nonumber \\
					&= \frac{E - V(\mathbf{R} + \mathbf{r}) + E^{~}_k(\mathbf{R} + \mathbf{r})}{
							4\pi\left( \bm{\mathfrak{R}}^2_{\mathrm{particle}} \right)
							\nu'_{\mathrm{vib}} \nu^{~}_{\mathrm{rot}}}.
			\end{align}
When a transition process takes place by emission process, the particle will evolve to a new state
			\begin{align}
				\label{65}
				m	&= \frac{E - E^{~}_0}{4\pi\left( \bm{\mathfrak{R}}^2_{\mathrm{particle}} \right)
										\nu'_{\mathrm{vib}} \nu^{~}_{\mathrm{rot}}} \nonumber \\
					&= \frac{E - V(\mathbf{R} + \mathbf{r}) + E^{~}_k(\mathbf{R} + \mathbf{r}) 
								- \Delta E^{~}_{\mathrm{emission}}}{4\pi\left( \bm{\mathfrak{R}}^2_{\mathrm{particle}} \right)
																	\nu'_{\mathrm{vib}} \nu^{~}_{\mathrm{rot}}} \nonumber \\
					&= \frac{E - V(\mathbf{R}' + \mathbf{r}') + E(\mathbf{R}' + \mathbf{r}')}{
								4\pi\left( \bm{\mathfrak{R}}^2_{\mathrm{particle}} \right)
								\nu'_{\mathrm{vib}} \nu^{~}_{\mathrm{rot}}}.					
			\end{align}
Eq. (\ref{65}) shows that, at every state, the mass of the particle will be always constant. However, mass is always constant at every state, but when a transition process takes place, there will be a difference in the perspective of mass between the new state and previous state. Eq. (\ref{65}) shows simply that if a transition process occurs, the change of the potential term is higher than the emission energy. Hence the kinetic energy must increase to a level higher than the previous state. We can express the relation of both $\mathbf{v}' + \mathbf{b}_*$ for $E'_{\mathrm{k}}$ (the total velocity at the new state) and $\mathbf{v} + \mathbf{b}$ for $E_k$ (total velocity at the previous state) as
			\begin{align}
				\label{66}
				\mathbf{v}' + \mathbf{b}_*
					= \left( \mathbf{v} + \mathbf{b} \right) + \Delta \mathbf{v}.
			\end{align}
We consider a case where the vector $\Delta \mathbf{v}$ perpendicular to the direction of the stationer velocity of quantum-sized volume, we have $\left( \mathbf{v} + \mathbf{b}\right) \perp \Delta \mathbf{v}$, so we can rewrite Eq. (\ref{65}) for a new state as
			\begin{align}
				\label{67}
				m 	= \frac{E - V(\mathbf{R}' + \mathbf{r}') + E^{~}_k(\mathbf{R} + \mathbf{r}) + \Delta E^{~}_k}{
							4\pi\left( \bm{\mathfrak{R}}^2_{\mathrm{particle}} \right)
							\nu''_{\mathrm{vib}} \nu^{~}_{\mathrm{rot}}}.
			\end{align}
Relative to the previous state, the mass of particle at previous state will be
			\begin{align}
				\label{68}
				m_0 	= \frac{E - V(\mathbf{R}' + \mathbf{r}') + E^{~}_k(\mathbf{R} + \mathbf{r}) }{
							4\pi\left( \bm{\mathfrak{R}}^2_{\mathrm{particle}} \right)
							\nu''_{\mathrm{vib}} \nu^{~}_{\mathrm{rot}}}.
			\end{align}
If we suppose that the time unit in every state must be represented by
			\begin{subequations}
				\begin{align}
					\label{69a}
					T = \frac{1}{\nu^{~}_{\mathrm{vib}}},
				\end{align}
then the internal time unit will be different for every state. In every transition process, a particle emitting or absorbing energy will change the internal time unit (proper time) where the change of the internal time unit must coincide with the emergence of the kinetic term of the particle (the difference of velocity between two states). Then we assume the internal time unit in one state will relativistically relate to another state, for example the internal time units at two states in Eq. (\ref{48}) and Eq. (\ref{49}) relate each other following
				\begin{align}
					\label{69b}
					T 	= \frac{1}{\nu^{~}_{\mathrm{vib}}}
						= \frac{1}{\nu''_{\mathrm{vib}}}
						= \frac{1}{\sqrt{1 - (\Delta \mathbf{v})^2/c^2}}
						= \frac{T''}{\sqrt{1 - (\Delta \mathbf{v})^2/c^2}}.
				\end{align}
			\end{subequations}

Furthermore we assume the perspective on mass between an occupied state and an unoccupied state or vice versa will also relativistically relate following
			\begin{align}
				\label{70}
				&\frac{E - V(\mathbf{R}' + \mathbf{r'}) + E_k(\mathbf{R} + \mathbf{r})}{
						4\pi\left( \bm{\mathfrak{R}}^2_{\mathrm{particle}} \right)
						\nu''_{\mathrm{vib}} \nu^{~}_{\mathrm{rot}}}
				\frac{1}{\sqrt{1 - (\Delta \mathbf{v})^2/c^2}} \nonumber \\
					&\qquad = m_0 \frac{1}{\sqrt{1 - (\Delta \mathbf{v})^2/c^2}} \nonumber \\
					&\qquad = \frac{E - V(\mathbf{R}' + \mathbf{r}') + E_k(\mathbf{R} + \mathbf{r}) + \Delta E_k}{								4\pi\left( \bm{\mathfrak{R}}^2_{\mathrm{particle}} \right)
						\nu''_{\mathrm{vib}} \nu^{~}_{\mathrm{rot}}} \nonumber \\
					&\qquad = m.
			\end{align}
Thus, for the case of the emission process, energy relating to the relativistically difference of mass between the two states must represent both magnetic energy generated by the emergence of the kinetic term (the difference of velocity) between two states and emitted energy (radiated energy) when the particle does a transition process, or two following equations must be the same
			\begin{subequations}
				\begin{align}
					\label{71a}
					\left(m - m^{~}_0\right)c^2
						&= m^{~}_0 \left[ \frac{1}{\sqrt{1 - (\Delta \mathbf{v})^2/c^2}} - 1\right] c^2 \nonumber \\
						&= \frac{1}{2}m^{~}_0 (\Delta \mathbf{v})^2 + \frac{3}{8}m^{~}_0\frac{(\Delta \mathbf{v})^4}{c^2}
							+ \frac{5}{16}m^{~}_0\frac{(\Delta \mathbf{v})^6}{c^4} \nonumber \\
						&\qquad 
							+ \frac{35}{128}m^{~}_0\frac{(\Delta \mathbf{v})^8}{c^6} + \cdots	,						
				\end{align}
				\begin{align}
					\label{71b}
					\left(m - m^{~}_0\right)c^2
						&= m^{~}_0\left[\frac{1}{\sqrt{1 - (\Delta \mathbf{v})^2/c^2}}\right] c^2 \nonumber \\
						&= E^{~}_{\mathrm{magnetic}} + \Delta E^{~}_{\mathrm{radiation}}.
				\end{align}
			\end{subequations}
			
For showing the equivalence of Eqs. (\ref{71a}) and (\ref{71b}), we use the wave function in Schr\"{o}dinger picture to describe the distribution of mass/charge/probability inside the quantum-sized volume, while the wave function in Heisenberg picture is used to describe the distribution of mass/charge/probability outside the quantum sized volume and relates to classical fields. We then connect the meaning of electrical-magnetic fields with the states of concentrated energy (in the term of particle's inclination to have minimum energy) in which the emergence of kinetic energy represents the increase of internal concentrated energy. Because the increase of internal concentrated energy serves the same effect as the decrease of the particle's tendency to come near to one another (the decrease of electric potential energy for hydrogen consisted of electron and nucleus is equivalent to magnetic energy in the entire space of outside of the quantum-sized volume.

Eq. (\ref{71a}) is the perspective of energy between two stationer states when the transition process takes place. Stationery current density at new state relatively to previous state is
			\begin{align}
				\label{72}
				\mathbf{J}(\mathbf{r}')
					&= \left[\left(\mathbf{v}' + \mathbf{b}^{~}_*\right) - \left(\mathbf{v} - \mathbf{b}\right)\right] \rho_e \\
				\label{73}
				\mathbf{J}(\mathbf{r}')
					&= \left[2 \mathbf{u} + \left(\mathbf{v}' - \mathbf{v}\right)\right]\rho_e,
			\end{align}
where $\mathbf{u} = \left(\mathbf{b}^{~}_* - \mathbf{b}\right)/2$. $\rho_e$ in Eqs. (\ref{72}) and (\ref{73}) satisfies Eq. (\ref{30})
			\begin{align}
				\label{74}
				\rho_e\left(\mathbf{r}'\right)
					= q\rho\left(\mathbf{r}'\right)
					= q\Psi^*(\mathbf{r}')\Psi(\mathbf{r}')
			\end{align}
with $q$ being the electron charge. Our model describes the random motion of the particle and allows us to see charge density and current density in Eqs. (\ref{73}) and (\ref{74}) as sources of classical electromagnetic field.
			\begin{align}
				\label{75}
				J^{\mu} = \left(c\rho, \mathbf{J}\right), \qquad r^{\mu} = \left(ct, \mathbf{r}\right).
			\end{align}						
Electromagnetic fields generated by these sources are
			\begin{align}
				\label{76}
				\partial_{\mu}F^{\mu\gamma}
					= \frac{4\pi}{c}J^{\gamma}; \qquad F^{\mu\gamma} = \partial^{\mu}A^{\gamma} - \partial^{\gamma}A^{\mu},
			\end{align}
			\begin{align}
				\label{77}
				A^{\mu} = \left(U, \mathbf{A}\right).
			\end{align}								
Working in Lorentz gauge \cite{MPD-2004} $\left(\partial^{\mu}A^{~}_{\mu} = 0\right)$, the classical result is
			\begin{align}
				\label{78}
				A^{\mu}\left(\mathbf{r}, t\right)
					= \frac{1}{c}
						\int \frac{1}{R}\, J^{~}_{\mu} \left(\mathbf{r}, t - \dfrac{R}{c}\right)\, \mathrm{d}^3 r'  \quad ;
						\quad R = \left|\mathbf{r} - \mathbf{r}'\right|,
			\end{align}						
where $R$ is the position vector of consideration of the magnetic field from new state. The magnetic field that corresponds to the vector potential $\mathbf{A}$ is
			\begin{align}
				\label{79}
				\mathbf{B}\left(\mathbf{r}, t\right)
					= \nabla \times \mathbf{A}
					= \nabla \times \frac{1}{c}\int \frac{1}{R}\,\mathbf{J} \left(\mathbf{r}', t - \dfrac{R}{c}\right)
						\,\mathrm{d}^3 r',
			\end{align}						
where unit vector of position is
			\begin{align}
				\label{80}
				\hat{\mathbf{n}}
					= \frac{\mathbf{r} - \mathbf{r}'}{\left|\mathbf{r} - \mathbf{r}\right|},
			\end{align}						
where $\mathbf{r}'$ is the position vector from the reference frame of the previous state to the reference frame of new state, and $\mathbf{r}$ is position vector from the reference frame of previous state to the point of consideration.

Using this vector we can rewrite Eq. (\ref{79}) as
			\begin{align}
				\label{81}
				\mathbf{B}\left(\mathbf{r},t\right)
					= \frac{1}{c}\int \hat{\mathbf{n}} \times \frac{\partial}{\partial R}
						\left[
							\left(\frac{1}{R}\right)\,\mathbf{J}\left(\mathbf{r}, t - \dfrac{R}{c}\right)
						\right]\,\mathrm{d}^3 r'.
			\end{align}						
For stationary current density, Eq. (\ref{81}) becomes
			\begin{align}
				\label{82}
				\mathbf{B}\left(\mathbf{r}, t\right)
					= \frac{1}{c}\int \hat{\mathbf{n}} \times 
						\left[-\frac{1}{R^2} \, J\left(\mathbf{r}', t - \frac{R}{c}\right)\right]\,\mathrm{d}^3 r'.
			\end{align}						
Applying Eq. (\ref{73}) to Eq. (\ref{83}), we have
			\begin{align}
				\label{83}
				\mathbf{B}\left(\mathbf{r}', t\right)
					= \frac{1}{c}\int \left(-\hat{\mathbf{n}}\right) \times
						\frac{1}{R^2}\left[2\mathbf{u} + \left(\mathbf{v}' - \mathbf{v}\right)\right] 
						\rho_e\left(\mathbf{r}'\right)\,\mathrm{d}^3 r'.
			\end{align}						
Using Eq. (\ref{74}) to Eq. (\ref{83}), we have
			\begin{align}
				\label{84}
				\mathbf{B}(\mathbf{r}, t)
					&= \frac{1}{c}\int \left(-\hat{\mathbf{n}}\right) \nonumber \\ 
					&\qquad\qquad \times
						\frac{q \Psi^*\left(\mathbf{r}'\right)\left[2\mathbf{u} + \left(\mathbf{v}' - \mathbf{v}\right)\right]
							\Psi\left(\mathbf{r}'\right)}{R^2}\,\mathrm{d}^3 r'.
			\end{align}						
The magnetic field generated by the new state from the previous state (\ref{84}) becomes
			\begin{align}
				\label{85}
				\mathbf{B}\left(\mathbf{r}, t\right)
					&= \frac{1}{c} \int \left(-\hat{\mathbf{n}}\right) \times
						\frac{q}{R^2}\Psi^*\left(\mathbf{r}'\right)(\Delta \mathbf{v})\Psi\left(\mathbf{r}'\right)
						\,\mathrm{d}^3 r' \\
				\label{86}
				\mathbf{B}\left(\mathbf{r}, t\right)
					&= -\frac{1}{c}\hat{\mathbf{n}} \times \frac{q \left\langle\Delta \mathbf{v}\right\rangle}{R^2},
			\end{align}
where $\mathbf{u} = \mathbf{b} - \mathbf{b}^{~}_*$ and $\Delta \mathbf{v}$ are as in Eq. (\ref{66}). Eq. (\ref{86}) is the Biot-Savart law showing that magnetic field is the non isotropic field shown by its dependence on the sinus angle. It is the same as magnetic field depending on the angle between the position vector of location of a short considered segment of wire and the current density vector, the perception of kinetic energy between the previous state and the new state (relative velocity) depends on the angle between both velocities (trajectories) in the two states. When we considered the relation between velocities in the previous and new states as Eq. (\ref{66})			
			\begin{align*}
				\mathbf{v}' + \mathbf{b}^{~}_*
					= \left(\mathbf{v} + \mathbf{b}\right) + \Delta \mathbf{v},
			\end{align*}
we have assumed that
			\begin{align*}
				\left(\mathbf{v} + \mathbf{b}\right) \perp \Delta \mathbf{v}.
			\end{align*}						
It means that from every point in the previous state, the particle in the new state always move with direction perpendicular to the position vector of location of a short segment of wire. On another word, the condition of $(\mathbf{v} + \mathbf{b}) \perp \Delta \mathbf{v}$ makes us consider the magnetic field as isotropic field.

Assume the density of magnetic fields is
			\begin{align}
				\label{87}
				U^{~}_{\mathrm{mag}}
					= \frac{E^{~}_{\mathrm{mag}}}{V}
					= \frac{\mathbf{B}^2}{2\mu^{~}_0}.
			\end{align}						
We can write Eq. (\ref{87}) as
			\begin{align}
				\label{88}
				\mathrm{d}E^{~}_{\mathrm{mag}}
					= \frac{\mathbf{B}^2}{2\mu^{~}_0}\,\mathrm{d}V.
			\end{align}						
Applying the magnetic field as an isotropic field and inserting Eq. (\ref{86}) into Eq. (\ref{88}), we have
			\begin{align}
				\label{89}
				\mathrm{d}E^{~}_{\mathrm{mag}}
					= \frac{q^2 \left\langle \Delta \mathbf{v}\right\rangle^2}{2c^2 \mu^{~}_0 R^4}\,\mathrm{d}V.
			\end{align}						
Using spherical coordinates and assuming the quantum-sized volume has a spherical shape with radius $r^{~}_{\mathrm{min}}$ so magnetic energy in the whole space is
			\begin{align}
				\label{90}
				E^{~}_{\mathrm{mag}}
					= \left[\frac{q^2 \left\langle \Delta \mathbf{v} \right\rangle^2}{2 c^2 \mu^{~}_0} \right]
						2 \pi \int\limits_0^{\pi} \sin \theta \,\mathrm{d}\theta 
						\int\limits_{r^{~}_{\mathrm{min}}}^{\infty} \frac{R^2}{R^4}\,\mathrm{d}R. 
			\end{align}						
We find
			\begin{align}
				\label{91}
				E^{~}_{\mathrm{mag}}
					= \left[\frac{2\pi q^2 \left\langle\Delta \mathbf{v}\right\rangle^2}{c^2 \mu^{~}_0}\right]
						\frac{1}{r^{~}_{\mathrm{min}}}.
			\end{align}						
Using relation $1/c = \mu^{~}_0/4\pi$, Eq. (\ref{91}) becomes
			\begin{align}
				\label{92}
				E^{~}_{\mathrm{mag}}
					= \left(\frac{\mu^{~}_0 q^2}{8\pi}\right) \frac{1}{r^{~}_{\mathrm{min}}} \left\langle\Delta \mathbf{v}\right\rangle^2.
			\end{align}						
Eq. (\ref{92}) must be equivalent to kinetic energy
			\begin{align*}
				E^{~}_{\mathrm{mag}}
					= \frac{1}{2}m^{~}_0\left\langle \Delta \mathbf{v}\right\rangle^2.
			\end{align*}						
So it should prevail that
			\begin{align}
				\label{93}
				m^{~}_{\mathrm{magnetic}}
					= \left(\frac{\mu^{~}_0 q^2}{4\pi}\right) \frac{1}{r_{\mathrm{min}}}
					= m_0.
			\end{align}						
Substitute the values of physical constant to Eq. (\ref{93}), where permeability in vacuum $(\mu^{~}_0 \,= \,4\,\pi\,\times10^{-7}\,$ $\mathrm{Wb/ A\cdot m})$, electron charge $(q = 1.602189 \times 10^{-19}\,\mathrm{C})$, and the chosen $r_{\mathrm{min}}$ is the classical radius of electron $\left(r^{~}_{\mathrm{min}} = 2.8179403 \times 10^{-15}\,\mathrm{m}\right)$, and we obtain:
			\begin{align}
				\label{94}
				m_{\mathrm{mag}}
					= \left(\frac{\mu_0 q^2}{4\pi}\right) \frac{1}{r_{\mathrm{min}}}
					= 9.10952 \times 10^{-31}\,\mathrm{kg}.
			\end{align}
Eq. (\ref{94}) shows that the magnetic mass is the same value as the rest mass, but this quantity is distributed to whole space of outside of the quantum-sized volume. The important result of this derivation is that classical radius of the electron must represent the radius of the quantum-sized volume. The particle that randomly move in the quantum-sized volume may be what Feynman refers to as a fuzzy ball \cite{RPF-RPL-MS-1963}.

Furthermore, we will show that the terms after the term $m_0 \left\langle \Delta \mathbf{v} \right\rangle^2/2$
			\begin{align}
				\label{95}
				\frac{3}{8}m^{~}_0 \frac{(\Delta \mathbf{v})^4}{c^2} + \frac{5}{16}m^{~}_0\frac{(\Delta \mathbf{v})^6}{c^2}
				+ \frac{35}{128}m^{~}_0\frac{(\Delta \mathbf{v})^8}{c^6} + \cdots ,
			\end{align}
must represent the transition energy from previous state to new state. If we assume that pure-energy or the total emitted-energy in any transition process should be the same (invariant) from every relative-rest frame and the total emitted energy must be equivalent to the generated-electromagnetic-field energy spreading all over space from every relative-rest frame. Fundamentally, existing electromagnetic fields that coincide with the transition process should represent the change of the tendency of particles to move away or towards each other and this can be represented by the change of both position and velocity of electron when the electron makes the transition. It will be equivalent to the change of electric and magnetic fields in time and it will be observed as electromagnetic field.

We consider the magnetic field from restively rest frame to both the previous and the new states. The magnetic field generated by the change of current when the transition process occurred, from Eq. (\ref{79}), is
			\begin{align}
				\label{96}
				\mathbf{B}\left(\mathbf{r}, t\right)
					= \frac{1}{c}\int \hat{\mathbf{n}} \times \left(-\frac{1}{Rc}\right)
							\mathbf{J}\left(\mathbf{r}', t - \frac{R}{c}\right)\,\mathrm{d}^3 r'.
			\end{align}
Following procedure and results by Davidson \cite{MPD-2004} having model from Schr\"{o}dinger picture to Heisenberg picture to accommodate the evolution of momentum operator in time
			\begin{align}
				\label{97}
				\int\Psi^*&\left(\mathbf{r}', t\right)\,\mathbf{a}\,\Psi\left(\mathbf{r}', t\right)\,\mathrm{d}^3 t
					\,\,\, \nonumber \\
					&\Rightarrow 
					\int\Psi^*\left(\mathbf{r}^{~}_{\mathrm{H}}, t\right)\, \mathbf{a}\,\Psi\left(\mathbf{r}^{~}_{\mathrm{H}}, t\right)
					\,\mathrm{d}^3 r^{~}_{\mathrm{H}},
			\end{align}							
where $\mathbf{r}^{~}_{\mathrm{H}}$ symbolizing representation position outside of the quantum-sized volume and considering the first order of Larmor's radiation, Davidson \cite{MPD-2004} finds that
			\begin{align}
				\label{98}
				\mathbf{B}\left(\mathbf{r},t\right)
					&= -\frac{q}{c^2 R^{~}_0} \hat{\mathbf{n}} \times
						\int \Psi^* \left(\mathbf{r}_{\mathrm{H}} , t\right) \mathbf{a}
							\Psi\left(\mathbf{r}_{\mathrm{H}}, t\right)\,\mathrm{d}^3 r \nonumber \\ 
					&= -\frac{q}{c^2 R^{~}_0} \hat{\mathbf{n}} \times \left\langle\Psi|\mathbf{a}|\Psi\right\rangle.
			\end{align}
In evaluating the radiation emitted, the limit where $R^{~}_0 = |\mathbf{r}| \rightarrow \infty$ is taken \cite{MPD-2004}. 

The transition of the electron from one state with kinetic terms to another state with certain other kinetic terms, accompanied by the change of the velocity of both quantum-sized volume and internal random velocities, from observer frame, can be considered as the movement of an electrical circuit or medium. So, by applying Faraday's law for the movement of a circuit, to the observer the electric field will be measured as \cite{RKW-1979}
			\begin{align}
				\label{99}
				\mathbf{E} = \mathbf{v}^{~}_e \times \mathbf{B},
			\end{align}							
where $\mathbf{v}^{~}_e$ is the velocity of a medium or circuit. The time interval of an electron for making transition process is $t^{~}_{\mathrm{transition}} = t^{~}_i + t^{~}_j$ where $t^{~}_i$ is the time interval needed by the electron to move from one point in the previous state to another point in the new state, and $t^{~}_j$ is the time interval needed by electron to complete a period of motion in the new state. $t^{~}_j$ consists of a motion period of the quantum-sized volume and or a period of internal random motion in new state. Thus, if $t^{~}_i$ and $t^{~}_j$ are the same order, electron making the transition process with interval time $t^{~}_{\mathrm{transition}}$ can be considered as the motion of circuit (wire) or medium with the time interval $t^{~}_{\mathrm{transition}}/2$ or with the velocity
			\begin{align}
				\label{100}
				\mathbf{v}^{~}_e \approx 2 \Delta \mathbf{v}.
			\end{align}							
In regions which are far from any charge or current, the relation between electric and magnetic fields in vacuum is \cite{MPD-2004}
			\begin{align}
				\label{101}
				\mathbf{E} = -\hat{\mathbf{n}} \times \mathbf{B}.
			\end{align}
Substitute Eqs. (\ref{100}) and (\ref{101}) into Eq. (\ref{98}), and we obtain Poynting vector
			\begin{align}
				\label{102}
				\mathbf{S}
					= \frac{c}{4 \pi} \mathbf{E} \times \mathbf{B}
					= \frac{q^2 |\Delta\mathbf{v}|}{2\pi c^3 R^2_0}
						\left\langle\Psi|\mathbf{a}|\Psi\right\rangle^2 \left(\sin^3 \theta\right) \hat{\mathbf{n}},
			\end{align}
where $\theta$ is the angle between $\left\langle\Psi|\mathbf{a}|\Psi\right\rangle$ and $\hat{\mathbf{n}}$ which is the same angle as between $\Delta\mathbf{v}$ and $\mathbf{B}$. The total radiated-power is
			\begin{align}
				\label{103}
				P_{\mathrm{rad}}
					= \int\limits_0^{\pi} \frac{q^2 |\Delta \mathbf{v}|}{2\pi c^3 R^2_0}
						\left\langle\Psi|\mathbf{a}|\Psi\right\rangle^2 \left(\sin^3 \theta\right)
						\left(2\pi R^2_0 \sin\theta\right)\,\mathrm{d}\theta.
			\end{align}
Computing Eq. (\ref{103}), we find
			\begin{align}
				\label{104}
				P^{~}_{\mathrm{rad}}
					= \frac{3}{8} \frac{q^2 |\Delta \mathbf{v}|}{c^3} \left\langle\Psi|\mathbf{a}|\Psi\right\rangle^2.
			\end{align}
To find the total radiated-energy, we integrate Eq. (\ref{104}) covering total transition time
			\begin{align}
				\label{105}
				E^{~}_{\mathrm{rad}}
					= \frac{3}{8} \frac{q^2 |\Delta \mathbf{v}|}{c^3}
						\int_0^{t^{~}_{\mathrm{transition}}}
						\left\langle\Psi|\mathbf{a}|\Psi\right\rangle^2 \,\mathrm{d}t.
			\end{align}
Since $t^{~}_{\mathrm{transition}} \approx r^{~}_{\mathrm{min}}/|\Delta \mathbf{v}|$, we can approximately approach the result of Eq. (\ref{105}) with
			\begin{align}
				\label{106}
				E^{~}_{\mathrm{rad}}
					&\approx \frac{3}{8} \frac{q^2 |\Delta \mathbf{v}|}{c^3} 
						\left\langle\Psi|\mathbf{a}|\Psi\right\rangle^2 
						\frac{r^{~}_{\mathrm{min}}}{|\Delta \mathbf{v}|} \nonumber \\
					&\approx \frac{3}{8} \frac{q^2 |\Delta \mathbf{v}|}{c^3}
						\left[|\Delta \mathbf{v}| \frac{|\Delta \mathbf{v}|}{r_{\mathrm{min}}}\right]^2
						\frac{r_{\mathrm{min}}}{|\Delta \mathbf{v}|} \nonumber \\
					&\approx \left(\frac{3}{8} \frac{q^2 |\Delta \mathbf{v}|^4}{c^3}\right) \frac{1}{r^{~}_{\mathrm{min}}}.
			\end{align}
Applying relation $1/c = \mu^{~}_0/4\pi$, we rewrite Eq. (\ref{106}) as
			\begin{align}
				\label{107}
				E^{~}_{\mathrm{rad}}
					\approx \frac{3}{8}\left(\frac{\mu^{~}_0 q^2}{4\pi} \frac{1}{r^{~}_{\mathrm{min}}}\right)
							\frac{(\Delta \mathbf{v})^4}{c^2}
					= \frac{3}{8} m_0 \frac{(\Delta \mathbf{v})^4}{c^2}.
			\end{align}
Eq. (\ref{107}) shows the first order of Larmor's radiation represents the second terms of
			\begin{align*}
				\left(m - m^{~}_0\right)c^2
					&= m^{~}_0 \left[\frac{1}{\sqrt{1 - (\Delta \mathbf{v})^2/c^2}} - 1\right] c^2 \nonumber \\
					&= \frac{1}{2}m^{~}_0 (\Delta \mathbf{v})^2 + \frac{3}{8}m^{~}_0\frac{(\Delta \mathbf{v})^4}{c^2}
						+ \frac{5}{16}m^{~}_0\frac{(\Delta \mathbf{v})^6}{c^4}  \nonumber \\
					&\qquad + \frac{35}{128}m^{~}_0 \frac{(\Delta \mathbf{v})^2}{c^6} + \cdots
			\end{align*}
This result proves that Eq. (\ref{71a}), besides representing the magnetic energy caused by the emergence of kinetic energy (the loss of the particle's tendency to approach each other is equivalent to the increase of the particle's tendency to keep away each other) must also represent the radiation energy when the transition process takes place between two states. The general form of magnetic field of the accelerated particle \cite{MPD-2004} is 
			\begin{align*}
				\mathbf{B}\left(\mathbf{r}', t\right)
					&= -\hat{\mathbf{n}} \times \frac{1}{c^2 R^{~}_0}
						\sum\limits_{p = 1}^{\infty}
						\int \frac{1}{(p-1)!} \nonumber \\
					&\qquad\qquad
							\left[\dfrac{\partial^p}{\partial t^p} \, \mathbf{J}
								\left(\mathbf{r}', t - \dfrac{R^{~}_0}{c}\right)
							\right]  
						\left[\dfrac{\hat{\mathbf{n}} \cdot \mathbf{r}'}{c}\right]^{p-1} \, \mathrm{d}^3 r'.
			\end{align*}
This equation can be rewritten with using electrical current observable as
			\begin{align}
				\label{108}
				\mathbf{B}\left(\mathbf{r}, t\right)
					= -\hat{\mathbf{n}} \times \frac{1}{c^2 R^{~}_0}
						\sum\limits_{p=1}^{\infty} \frac{1}{(p-1)!} \frac{\partial^p}{\partial t^p_0} \,
						\mathbf{I}^{~}_p\left(t^{~}_0\right)\left(\frac{1}{c}\right)^{p-1},
			\end{align}
where 
			\begin{align}
				\label{109}
				\mathbf{I}_p \left(t_0\right)
					= \int \mathbf{J} \left(\mathbf{r}', t_0\right) \left(\frac{\hat{\mathbf{n}} \cdot \mathbf{r}'}{c}\right)^{p-1}
					\,\mathrm{d}^3 r'; \qquad t_0 = t - \frac{R^{~}_0}{c}.
			\end{align}
Doing transformation into Heisenberg picture, Davidson \cite{MPD-2004} found
			\begin{align}
				\label{110}
				\mathbf{I}_p \left(t^{~}_0\right)
					&= \left[\frac{q}{2M} 
						\int\Psi^*\left(\mathbf{r}^{~}_{\mathrm{H}}, 0\right)  
						\left\lbrace
							\mathbf{P}^{~}_u\left(t^{~}_0\right) \left[\hat{\mathbf{n}} 
								\cdot \mathbf{R}^{~}_v\left(t^{~}_0\right)\right]^{p-1}\right.\right.\nonumber \\
							 	&\qquad  + \left.\left.\left[
								\hat{\mathbf{n}} \cdot \mathbf{R}^{~}_{v}\left(t^{~}_0\right)
							\right]^{p-1} \mathbf{P}^{~}_u\left(t^{~}_0\right)
						\right\rbrace
						\Psi\left(\mathbf{r}^{~}_{\mathrm{H}}, 0 \right) \, \mathrm{d}^3 r^{~}_{\mathrm{H}}
						\right.\bm{\biggl ]} \nonumber \\
					&\qquad + \left[\frac{q}{M} 
						\int\Psi^* \left(\mathbf{r}^{~}_{\mathrm{H}}, 0\right)
							\left\lbrace
								P_{qv}\left[
										\hat{\mathbf{n}} \cdot \mathbf{R}^{~}_v\left(t^{~}_0\right)
										\right]^{p-1}
							\right\rbrace\right. \nonumber \\
							&\qquad\qquad\Psi\left(\mathbf{r}^{~}_{\mathrm{H}}, 0 \right)\,\mathrm{d}^3 r^{~}_{\mathrm{H}}
						\bm{\bigg ]}.
			\end{align}							
Linking Eq. (\ref{110}) to the calculation of the second term of Eq. (\ref{86}), we can conclude that the other terms
			\begin{align*}
				\frac{5}{16}m^{~}_0\frac{(\Delta \mathbf{v})^6}{c^4} 
						+ \frac{35}{128}m^{~}_0 \frac{(\Delta \mathbf{v})^2}{c^6} + \cdots
			\end{align*}
must be served by
			\begin{align}
				\label{111}
				\mathbf{B}\left(\mathbf{r}', t\right)
					&= -\hat{\mathbf{n}} \times \frac{1}{c^2 R^{~}_0}
						\sum\limits_{p = 1}^{\infty}
						\int \frac{1}{(p-1)!}  \nonumber \\
					&\qquad\qquad
							\left[\dfrac{\partial^p}{\partial t^p} \, \mathbf{J}
								\left(\mathbf{r}', t - \dfrac{R^{~}_0}{c}\right)
							\right]
					\left[\dfrac{\hat{\mathbf{n}} \cdot \mathbf{r}'}{c}\right]^{p-1} \, \mathrm{d}^3 r',
			\end{align}
with $p > 1$. From applying the invariance of radiated energy, our proposal indicates that the quantization of a classical wave must represent the scheme of second quantization.

		\subsection{\label{sec:5-5}Radiated-energy in spin transition}
		
Eq. (\ref{71a}) is the general term of radiation for the transition process including the change of both internal random motion and the quantum-sized volume velocities. If we only see the transition when the velocity of the quantum-sized volume does not change or the transition process only represents spin-state transition (the change of internal random motion), we get that magnetic and radiated-energy of transition process still keep the form
			\begin{align}
				\label{112}
				\left(m - m^{~}_0\right)
					&= m^{~}_0 \left(1 - \frac{1}{\sqrt{1 - (\mathbf{b}_* - \mathbf{b})^2/c^2}}\right) c^2 \nonumber \\
					&= \frac{1}{2} m^{~}_0 \left(\mathbf{b}^{~}_* - \mathbf{b}\right)^2 
						+ \frac{3}{8} m^{~}_0 \frac{\left(\mathbf{b}^{~}_* - \mathbf{b}\right)}{c^2} \nonumber \\
					&\qquad	+ \frac{5}{16}m^{~}_0\frac{\left(\mathbf{b}^{~}_* - \mathbf{b}\right)}{c^4} + \cdots,
			\end{align}					
where it still prevails that
			\begin{align}
				\label{113}
				E^{~}_{\mathrm{mag}}
					= \frac{1}{2}m^{~}_0 \left\langle\mathbf{b}_* - \mathbf{b}\right\rangle^2
					= \left(\frac{\mu^{~}_0 q^2}{8 \pi}\right) \frac{1}{r^{~}_{\mathrm{min}}}
						\left\langle\mathbf{b}^{~}_* - \mathbf{b}\right\rangle.
			\end{align}					
with $r^{~}_{\mathrm{min}}$ is the classical radius of electron, and
			\begin{align}
				\label{114}
				E^{~}_{\mathrm{rad}}
					\approx \frac{3}{8} \left(\frac{\mu^{~}_0 q^2}{4 \pi} \frac{1}{r_{\mathrm{min}}}\right)^{~}_{\mathrm{rad}}
							\frac{\left\langle\mathbf{b}^{~}_* - \mathbf{b}\right\rangle^4}{c^2}
					= \frac{3}{8} m^{~}_0 \frac{\left\langle \mathbf{b}^{~}_* - \mathbf{b} \right\rangle^4}{c^2}.
			\end{align}					
This prevails for firs order of radiated-energy.
		
		\subsection{\label{sec:5-6}Spin-spin and spin-orbit interactions}
			
Here, we discuss the interaction potentials that accompany the process of transition between two states. We consider the case in which an emission process is accompanied by the emergence of internal spinning motion when the transition process that is taking place coincides with internal diffusion process. Spin or internal spinning motion is shown by the emergence of velocity
			\begin{align*}
				\mathbf{u}
					= \frac{\hslash}{2m}\frac{\nabla \rho}{\rho}.
			\end{align*}
As in Eq. (\ref{38}), this velocity describes the change of the velocity of internal motion in the quantum-sized volume when a transition process takes place. We see interaction potentials accompanying the transition process for the kinds of interaction in which the velocity of the quantum-sized volume does not change. There are two kinds of interactions for this transition. One is the energy-emitting process that makes the random internal velocity decrease (positive diffusion) and the other is the energy-absorbing 	process that makes the random internal velocity increase.
			
From Eq. (\ref{72}), relative to the nucleus, the current density of an electron is
			\begin{align*}
				\mathbf{J}^{~}_0 \left(\mathbf{r}', t\right)
					= \left(\mathbf{b} + \mathbf{v}\right) \rho_e \left(\mathbf{r}', t\right).
			\end{align*}
This current will generate the magnetic field
			\begin{align}
				\label{115}
				\mathbf{B}\left(\mathbf{r}, t\right)
					&= \nabla \times \mathbf{A} \nonumber \\
					&= \nabla \times \frac{1}{c}
						\int \frac{1}{R}\,\mathbf{J}^{~}_0 \left(\mathbf{r}', t - \dfrac{R}{c}\right)
						\,\mathrm{d}^3 r'.
			\end{align}
Insert the current density into Eq. (\ref{115}), and we have
			\begin{align}
				\label{116}
				\mathbf{B}\left(\mathbf{r}, t\right) 
					&= \nabla \times \mathbf{A} \nonumber \\
					&= \frac{1}{c} 
						\int \frac{1}{R}\,\left\lbrace 
									\nabla \rho_e
										\left(
											\mathbf{r}' , t - \dfrac{R}{c}
										\right) \times
										\left(
											\mathbf{b} + \mathbf{v}
										\right) \right. \nonumber \\
										&\qquad +
										\rho_e
										\left.\left(
											\mathbf{r}', t - \dfrac{R}{c}
										\right)
										\left[
											\nabla \times
											\left(
												\mathbf{b} + \mathbf{v}
											\right)
										\right]
								\right\rbrace\, \mathrm{d}^3 r'.
			\end{align}
We may rewrite Eq. (\ref{116}) as
			\begin{align}
				\label{117}
				\mathbf{B}\left(\mathbf{r}, t\right) 
					&= \nabla \times \mathbf{A} \nonumber \\
					& = \frac{1}{c}\int
						\frac{1}{R}\,
							\left[\nabla \rho_e \left(\mathbf{r}', t - \dfrac{R}{c}\right) \times \left(\mathbf{b} 
						+ \mathbf{v}\right)\right. \nonumber \\ 
						&\qquad + \rho_e\left.\left(\mathbf{r}', t 
						- \dfrac{R}{c}\right)\hat{\mathbf{n}} \times \frac{\partial }{\partial R} \left(\mathbf{b} 
						+ \mathbf{v}\right)\right]\,\mathrm{d}^3 r'.
			\end{align}

If the electron emits energy (i.e. it undergoes a transition process) and its velocities $(\mathbf{b}, \mathbf{v})$ do not change in relation to position $R$, so the magnetic field generated by the change of the distribution of charge density $\rho_e$ is
			\begin{align}
				\label{118}
				\mathbf{B}\left(\mathbf{r}, t\right)
					&= \nabla \times \mathbf{A} \nonumber \\
					&= \frac{1}{c} \int
						\frac{1}{R}\,\nabla \rho_e \left( \mathbf{r}' , t - \dfrac{R}{c}\right) 
							\times \left(\mathbf{b} + \mathbf{v}\right)\,\mathrm{d}^3 r'.
			\end{align}
Eq. (\ref{118}) may be written as
			\begin{align}
				\label{119}
				\mathbf{B}\left(\mathbf{r}, t\right)
					&= \nabla  \times \mathbf{A} \nonumber \\
					&= \frac{1}{c^2}
						\int \frac{1}{R}\,\dot{\rho}_e\left(\mathbf{r}, t - \dfrac{R}{C}\right) \times \left(\mathbf{b} + \mathbf{v}\right)\,\mathrm{d}^3 r'.
			\end{align}
If we only focus on the diffusion process that relates to the change of internal random motion (the density of diffusion current), with using (\ref{35}), Eq. (\ref{119}) becomes
			\begin{align}
				\label{120}
				\mathbf{B}\left(\mathbf{r}, t\right)
					&= \nabla \times \mathbf{A} \nonumber \\
					&= \frac{1}{c^2}
						\int\frac{1}{R}\,\left\lbrace
									\frac{\hslash}{2m}\nabla^2
										\left[
											\rho_e \left(\mathbf{r}', t - \dfrac{R}{c}\right)
										\right]
									\right\rbrace \nonumber \\
					&\qquad\qquad \times \left(\mathbf{b} + \mathbf{v}\right)\,\mathrm{d}^3 r'.
			\end{align}
Recall Eq. (\ref{32}) and (\ref{33})
			\begin{align}
				\label{121}
				\rho_m
					= m \Psi^*\left(\mathbf{r}', t\right) \Psi\left(\mathbf{r}', t\right); \qquad
				\rho_e
					= q \Psi^*\left(\mathbf{r}',t\right) \Psi\left(\mathbf{r}', t\right).
			\end{align}
Applying (\ref{121}) into Eq. (\ref{120}), we obtain
			\begin{align}
				\label{122}
				\mathbf{B}\left(\mathbf{r}, t\right)
					&= \nabla \times \mathbf{A} \nonumber \\
					&= \frac{q}{mc^2}
						\int \frac{1}{R}\,\left\lbrace
										\frac{\hslash}{2m}\nabla^2
										\left[
											\rho_m \left(\mathbf{r}', t - \dfrac{R}{c}\right)
										\right]
									\right\rbrace \nonumber \\
					&\qquad\qquad \times \left(\mathbf{b} + \mathbf{v}\right)\,\mathrm{d}^3 r'.
			\end{align}						
Inserting $\mathbf{u} = \left(\hslash/2m\right)\left(\nabla \rho \right)/\rho$ into Eq. (\ref{122}), we obtain
			\begin{align}
				\label{123}
				\mathbf{B}\left(\mathbf{r}, t\right)
					&= \nabla \times \mathbf{A} \nonumber \\
					&= \frac{q}{mc^2}
						\int\frac{1}{R}\,\left\lbrace
									\nabla \cdot \left[\mathbf{u}\rho_m
										\left(\mathbf{r}', t - \dfrac{R}{c}\right)
									\right]\right\rbrace \nonumber \\
					&\qquad\qquad	\times \left(\mathbf{b} + \mathbf{v}\right)\,\mathrm{d}^3 r'.
			\end{align}
If the distribution of the $\mathbf{u}$ field does not relatively change to the internal position, Eq. (\ref{123}) may be rewritten by
			\begin{align}
				\label{124}
				\mathbf{B}\left(\mathbf{r}, t\right)
					&= \nabla \times \mathbf{A} \nonumber \\
					&= \frac{q}{mc^2}
						\int\frac{1}{R}\left\lbrace
										\mathbf{u}\cdot \left[\nabla\rho_m
										\left(\mathbf{r}', t - \dfrac{R}{c}\right)
									\right]\right\rbrace \nonumber \\
					&\qquad\qquad \times\left(\mathbf{b} + \mathbf{v}\right)\,\mathrm{d}^3 r'.
			\end{align}
If we move from the Schr\"{o}dinger picture to Heisenberg picture and ignore the distribution of mass in space, Eq. (\ref{124}) may be written by
			\begin{align}
				\label{125}
				\mathbf{B}\left(\mathbf{r}, t\right)
					&= \nabla \times \mathbf{A} \nonumber \\
					&= \frac{q}{mc^2}
						\frac{h}{2}\frac{\nabla \rho}{\rho}
						\int\frac{1}{R}\,\left[
									\nabla\Psi^*\left(\mathbf{r}^{~}_{\mathrm{H}}, t\right)
									\Psi\left(\mathbf{r}^{~}_{\mathrm{H}},t\right)
									\right] \nonumber \\
					&\qquad\qquad\qquad \times \left(\mathbf{b} + \mathbf{v}\right)\,\mathrm{d}^3 r^{~}_{\mathrm{H}}.
			\end{align}	
Furthermore, we propose that
			\begin{align}
				\label{126}
				\int\nabla \Psi^*\left(\mathbf{r}^{~}_{\mathrm{H}},t\right) \Psi\left(\mathbf{r}^{~}_{\mathrm{H}}, t\right)
				\,\mathrm{d}^3 r^{~}_{\mathrm{H}}
					= \frac{1}{4\pi \epsilon_0 R^2}.
			\end{align}	
So we obtain 
			\begin{align}
				\label{127}
				\mathbf{B}\left(\mathbf{r}, t\right)
					&= \nabla \times \mathbf{A} \nonumber \\
					&= \frac{q}{mc^2}
						\frac{1}{4\pi \epsilon^{~}_0 R^3}\left(\dfrac{h}{2} \dfrac{\nabla \rho}{\rho}\right) \times \left(\mathbf{b} + \mathbf{v}\right).
			\end{align}	
From electron frame, the kinetic energy of both the nucleus and the observer will relatively increase because the emitting of energy makes its internal kinetic energy (internal motion velocity in the quantum-sized volume) decrease. The increase of the kinetic energy of nucleus or observer from electron frame will responded by electron with the loss of its tendency to come closer to nucleus where the loss of this tendency will represent the electron in magnetic field. By contrast, the nucleus and observer will see the decrease of the kinetic energy of electron as the increase of the particle's tendency to get closer to the nucleus. Another possibility is that the nucleus and observer will see Eq. (\ref{127}) as electric field
			\begin{align}
				\label{128}
				\mathbf{E}\left(\mathbf{r}, t\right)
					= \frac{q}{mc^2}
						\frac{1}{4\pi \epsilon^{~}_0 R^3}\left(\dfrac{h}{2} \dfrac{\nabla \rho}{\rho}\right) \times \left(\mathbf{b} + \mathbf{v}\right).
			\end{align}
At the nucleus frame, interaction potential between the electron and nucleus will be the same as work
			\begin{align}
				\label{129}
				V^{\mathrm{interaction}}
					&= - \int_\infty^R \mathbf{F}^{~}_c \cdot \mathrm{d}\mathbf{r} \nonumber \\
					&= \frac{q^2}{4 \pi \epsilon^{~}_0 mc^2}
						\left[
							\left(\frac{h}{2}\frac{\nabla \rho}{\rho}\right) 
							\times \left(\mathbf{b} + \mathbf{v}\right)
						\right] \cdot
						\int_{\infty}^R \frac{\mathrm{d}\mathbf{r}}{R^3}  \nonumber \\
					&= -\frac{q}{mc^2} \left[
											\left(
												\frac{h}{2}\frac{\nabla \rho}{\rho}\right) \times
												\left(\mathbf{b} + \mathbf{v}\right)
										\right] \cdot
						\frac{1}{4\pi \epsilon^{~}_0} \frac{q}{R^2}\,\hat{\mathbf{r}}.
			\end{align}		
Eq. (\ref{129}) can be rewritten by
			\begin{align}
				\label{130}
				V^{\mathrm{so}}
					= -\frac{q}{mc^2}\left[
										\left(\frac{h}{2} \frac{\nabla\rho}{\rho}\right)
										\times
										\left(\mathbf{b} + \mathbf{v}\right)
									\right] \cdot \mathbf{E}
			\end{align}	
We choose $\left(h/2\right) \left(\nabla \rho\right)/\rho = \mathbf{P}^{~}_S$ to be the internal momentum that emerges when any transition process takes place and connects to the transition velocity $\mathbf{u}$. Eq. (\ref{130}) may be rewritten by
			\begin{align}
				\label{131}
				V^{\mathrm{interaction}}
					&= -\frac{q}{mc^2}
						\left[
							\mathbf{P}^{~}_S \times \left(\mathbf{b} + \mathbf{v}\right)
						\right] \cdot \mathbf{E}  \nonumber \\
					&= -\frac{q}{mc^2}
							\mathbf{P}^{~}_S \cdot \left[\left(\mathbf{b} + \mathbf{v}\right)
						 \times \mathbf{E} \right].
			\end{align}							
Eq. (\ref{131}) expresses Biot-Savart law
			\begin{align}
				\label{132}
				\mathbf{B}
					&= -\frac{\left[\left(\mathbf{b} + \mathbf{v}\right) \times \mathbf{E}\right]}{c} \nonumber \\
					&= -\left[
							\frac{\left(\mathbf{v} \times \mathbf{E}\right)}{c} + 
							\frac{\left(\mathbf{b} \times \mathbf{E}\right)}{c}
						\right] \nonumber \\
					&= \mathbf{B}^{~}_0 + \mathbf{B}^{~}_s,
			\end{align}									
where $\mathbf{B}^{~}_0$ is the magnetic field generated by orbital motion, and $\mathbf{B}^{~}_s$ is the magnetic field generated by internal motion. We re-express Eq. (\ref{131}) as
			\begin{align}
				\label{133}
				V^{\mathrm{interaction}}
					= V^{\mathrm{so}} + V^{\mathrm{ss}}
					= \frac{q}{mc} \mathbf{P}^{~}_S \cdot \mathbf{B}^{~}_0 + 
						\frac{q}{mc} \mathbf{P}^{~}_S \cdot \mathbf{B}^{~}_s,
			\end{align}														
where $V^{\mathrm{so}}$ and $V^{\mathrm{ss}}$ respectively are potentials for spin-orbit and spin-spin interactions.

		\subsection{\label{sec:5-7}The electron in external magnetic field}

In our model has proposed that before the external treatment, electrons have been at stationer state with internal potential $V$. When an external field is presen	t, it will make an electron undergo the transition process demonstrated by the equation
			\begin{align*}
				\frac{1}{2}\left(D D^{~}_* + D^{~}_* D\right)\mathbf{x}(t)
					&= \mathbf{a}(t) \nonumber \\
					&= -\frac{1}{m}\nabla \left(V + V^{~}_{\mathrm{external}}\right) \nonumber \\
					&= \frac{1}{m} \left(\mathbf{F}^{~}_{\mathrm{system}} + \mathbf{F}^{~}_{\mathrm{external}}\right)
			\end{align*}										
If the electron has been in a physical system with the potential ($e\mathbf{E}$) such as in a Coulomb potential with a nucleus, when one applies an external magnetic field the electron will obey motion equations such as Eqs. (\ref{39}) and (\ref{40}) that are
			\begin{align}
				\label{134}
				\frac{\partial \mathbf{u}}{\partial t}
					= -\beta\nabla \left[\nabla \cdot \left(\bm{\upsilon} + \mathbf{v}\right)\right] - 
						\nabla\left[\mathbf{u} \cdot \left(\bm{\upsilon} + \mathbf{v}\right) \right]
			\end{align}										
			\begin{align}
				\label{135}
				\frac{\partial (\bm{\upsilon} + \mathbf{v})}{\partial t}
					&= \frac{e}{m} \left[
									\mathbf{E} + \left(\frac{1}{c}\right)
									\left(\bm{\upsilon} + \mathbf{v}\right) \times \mathbf{H}
								\right] 
									+ (\mathbf{u} \cdot \nabla)\mathbf{u}\nonumber \\
					&\qquad\qquad	+ \frac{\hslash}{2m} \nabla \mathbf{u}
									- \left(\bm{\upsilon} \cdot \nabla\right)
										\left(\bm{\upsilon} + \mathbf{v}\right),
			\end{align}
where
			\begin{align}
				\label{136}
				\mathbf{F}^{~}_{\mathrm{external}}
					= \mathbf{F}^{~}_{\mathrm{magnetic}}
					= \left(\frac{1}{c}\right)\left(\bm{\upsilon + \mathbf{v}}\right) \times \mathbf{H}.
			\end{align}										
Similar to the time-dependent Schr\"{o}dinger equation, there are two general cases for this treatment. For the case in which $\mathbf{v} = 0$, Eqs. (\ref{134}) and (\ref{135}) are equivalent to the Schr\"{o}dinger equation
			\begin{align}
				\label{137}
				i\hslash \frac{\partial \Psi}{\partial t}
					= \frac{1}{2m} \left(-i \hslash \nabla +
										\frac{e}{c}\mathbf{A}
									\right)^2 \Psi + e \phi \Psi,
			\end{align}										
with $\Psi = e^{R + iS}$ and $S$ fulfill $\nabla S = \left(m/\hslash\right) \left[\bm{\upsilon} + e\mathbf{A}/mc \right]$. Whereas for the cases where $\mathbf{v} = 0$, Eqs. (\ref{134}) and (\ref{135}) are equivalent to the modified-Schr\"{o}dinger equation
			\begin{align}
				\label{138}
				i\hslash \frac{\partial \Psi}{\partial t}
					&= \frac{1}{2m} \left(-i \hslash \nabla + \frac{e}{c} \mathbf{A} \right)^2 \Psi \nonumber \\
					&\qquad + 
						\left[e \phi + E^{~}_k
						\int m \left(\bm{\upsilon} \nabla \cdot \mathbf{v}\right)\, \mathrm{d}^3 r'\right] \Psi,
			\end{align}			 						
where $E^{~}_k = mv^2/2$, $\Psi = e^{R + iS}$ with $\nabla S = \left(m/\hslash\right)$ $\left[ \left(\bm{\upsilon} + \mathbf{v}\right) + \left(e/mc\right) \mathbf{A}\right]$ and $E^{~}_k = mv^2/2$. For a special case when $\mathbf{u}$ is a solenoidal vector field ($\nabla \cdot \mathbf{v} = 0$) then the Eq. (\ref{138}) becomes
			\begin{align}
				\label{140}
				i\hslash \frac{\partial \Psi}{\partial t}
					= \frac{1}{2m} \left(-i\hslash \nabla + \frac{e}{c}\mathbf{A}\right)^2\Psi
						\left(e \phi + E^{~}_k\right)\Psi
			\end{align}

	\section{\label{sec:6}DISCUSSION}

The main point of the development of our model is that the displacement of the particle in a non-stationery case not only is contributed by both the diffusion process and the velocity field of internal random motion as classical Brownian motion
		\begin{align*}
			\mathrm{d}\mathbf{x}(t)
				= \mathbf{b}
					\bm{(}
						\mathbf{x}(t), t
					\bm{)} +
					\mathrm{d}\mathbf{W}(t),
		\end{align*}
where $\mathbf{W}(t)$ is the Wiener process, but also is contributed to by the movement of the quantum-sized volume. As proposed by Recami and Salesi \cite{ER-GS-1998}, the movement of the quantum-sized volume is a "classical" part of motion while the internal random motion is the "quantum" part of motion. The internal random motion only changes when the transition processes (internal diffusion processes) occur where these processes are always accompanied by emitting or absorbing energies and will generate spinning motion in the quantum-sized volume. The internal random velocity field has maximum value at the speed of light when there is an interaction perspective (with the physical system or environment), which ensures that there are no nodal surface for every stationary state. In Nelson's work \cite{EN-1966}, current velocity will be generated only by the diffusion process, so it makes the increase of current velocity strongly determine the smoothness of $\bm{\upsilon}$, $\mathbf{u}$, and $\rho$ (i.e. how the velocities spread in space). Additionally it indicates that the model faces difficulty in covering the transition process at a high velocity. The motion of the quantum-sized volume coupled with the change of internal mass/charge/probability density is actually posed in order to ensure the velocity fields ($\bm{\upsilon}$ and $\mathbf{u}$) and the density probability ($\rho$) become the smooth functions for both the stationer and the transition process. This approach is totally different from conventional Brownian motion models, which view the displacement of the particle for time-dependent cases and this is only generated by the diffusion process and velocity of Brownian motion.

In our model, we also pose that random motion is limited in the quantum-sized volume. This is in order to support the realistic explanation of quantum mechanics that has to meet at least one of three requirements of the charge of a particle concentrated in a small volume of space as stated by Jung \cite{KJ-2009}. The probability density of finding the particle in the quantum-sized volume will change due to the transition process where it will be determined by the changes of the quantum-sized volume, random motion and diffusion velocities. However the total probability density of all space in the quantum-sized volume (the probability of finding the particle in the quantum-sized volume) as always one. Furthermore, because we can interchange probability density with charge and mass densities, the fact that the total number of probability density of finding the particle in the quantum-sized volume is always the same makes the total of mass and charge in the quantum-sized volume is always constant ($m$ and $q$), whatever the mass and charge density distributed in the quantum-sized volume. This property strongly supports our concepts on mass in that it must be invariant in every state.

Mass, quantity, even though it is always constant at every state in both stationary and transition processes, can be different if we consider two states when the particle evolves from one stationary state to another. A new state occupied by a transited-electron with higher kinetic energy will get the rest-mass perspective relative to its left-state. The energy being equal to the difference of mass between previous state (the state that is left by the particle) and new state (the state is occupied by the particle) is
		\begin{align*}
			\left(m - m^{~}_0\right) c^2
				&= m^{~}_0 \left(\frac{1}{\sqrt{1 - (\Delta \mathbf{v})^2/c^2}}\right) c^2 \\
				&= \frac{1}{2} m^{~}_0(\Delta \mathbf{v})^2
					+ \frac{3}{8} m^{~}_0 \frac{\left(\Delta \mathbf{v}\right)^4}{c^2}
					+ \frac{5}{16} m^{~}_0 \frac{\left(\Delta \mathbf{v}\right)^6}{c^4} \nonumber \\
				&\qquad + \frac{35}{128} m^{~}_0 \frac{\left(\Delta \mathbf{v}\right)^8}{c^6} + \cdots
		\end{align*}		 							
This energy, as shown above, will represent the magnetic energy that is equivalent with kinetic energy and the transition energy (emission and absorption energies) between a left state and an occupied state.
		\begin{align*}
			\left(m - m^{~}_0\right) c^2
				&= m^{~}_0 \left(\frac{1}{\sqrt{1 - (\Delta \mathbf{v})^2/c^2}} - 1\right) c^2 \nonumber \\
				&= E^{~}_{\mathrm{magnetic}} + E^{~}_{\mathrm{transition}}
		\end{align*}
This result shows that relativistic effects such as related to time interval should not be enough to be considered in the framework of kinematic aspects as established-understanding claimed by Einstein. Our interpretation shows that internal random motion exhibiting random vibrating system and determining internal time interval unit (proper time) for states must play a fundamental role in relativistic effects and it can depict the change of both potential and kinetic terms as well as the transition energy of the particle. Furthermore, from the seeking relation between mass and electric-magnetic fields, we have posed its tendency to attract particles of the same charges or repel different charges. Based on this viewpoint, we can understand the magnetic energy (magnetic field) as the lost of the particle's tendency for having minimum concentrated-energy $E$	caused by the increase of the particle's velocity (the kinetic energy of particles).	
		
We have also shown that Eq. (\ref{71a}) still prevail to picture the state transition of spin representing the increase or decrease of internal velocity ($\mathbf{b}$ or $\mathbf{b}_*$) when diffusion process take places. The relativistic effect on to the mass when the transition process takes place must be generated by unit times relativistically changing by the emergence of the kinetic aspect. We also showed that spin-spin and spin-orbit interactions (for the case of hydrogen atom) must be caused by the difference perspective on kinetic energy between the electron and its nucleus. The increase of kinetic energy (the increase of magnetic field) that is felt by one particle can be felt as the decrease of kinetic energy (the increase of electric field) by another particle.

In addition to the proposal of a new basic-fundamental connection between mass and electromagnetic fields, we also found that if we view absorbed-radiated energy in terms of light, our model indicates that electromagnetic fields should be not directly represented by the absorbed-radiated energy. Instead, electromagnetic fields should just represent the physical effects accompanying the process of absorbed-radiated energy generated by the transition process (or, how environment responds to the change of the transition process) where electromagnetic field can be used to determine the equivalence value of the amount of absorbed-radiated energy. We propose that if the electromagnetic wave must by the representation of wave properties of light, so must the pure radiated-absorbed energy represent the particle properties of the light. Furthermore, because the particle is modeled as energy localized on the surface of dimensional sphere-form (2-manifold without boundary) we can imagine that our real space may be composed of the space-particles sea where pure energy can be localized or propagated with particle properties.

The mass feature of particles in our model is the physical condition generated by concentrated-energy located in the surface of 3-dimensional sphere-form. Using this viewpoint, we can understand the environment or the existence of system as the condition accompanying the loss of concentrated-energy. Thus, the total lost concentrated-energy from the particle must be compensated by the emergence of potentials and the kinetics terms as well as the change of velocity of internal random motion. The change of the concentrated-energy of the particle will describe the dynamics of the system relative to the particle or the dynamics of the particle relative to the system. Therefore the energy in Schr\"{o}dinger Eq. (\ref{44}) and (\ref{45}) must represented the total energy radiated by the particle since there is no interaction perspective. We can also see that for the case of simple like an atom, the quantization of emitted absorbed energy will relate to states quantified by the potential and kinetic terms.

The principle difference between our model and previous stochastic models lies in the causality of the emergence of random motion. In our model, we propose motion must be generated by concentrated-energy localized on the surface of a particle. Therefore, we do not only need quantum potential (external potential) to generate internal motion but also a bath (ether or vacuum fluctuation or a noise field background as the main source of fluctuation) to generated random motion and diffusion. In our proposal concepts on mass, we view the background noise field [Eq. (\ref{11})] that depicts environmental fluctuation must exist to interact with a stationer system. The interactions do not change on average, the energy of the stationary system and only affect the probabilistic character of outcome.

In the context of the debate on how wave function evolves by the Schr\"{o}dinger equation in a predictable deterministic way, despite when a physical quantity is measured, the outcome is not predictable in advance \cite{SLA-AB-2009}, our model demonstrate that the quantum-sized volume "bringing" the probability density will evolve and coincide with transition processes in a predictable way. But, the information of physical quantity coupled with position (direction) is randomly stored in the quantum-sized volume, and because every measurement is a process to cause transition between states where this process is accompanied by the change of probability density to find the particle in the quantum-sized volume, thus, the measurement will determine the behavior of statistical outcome for every physical observable coupled to (defined) position (direction). However the statistical outcome can also be generated by a background noise field or environment fluctuation and it does not affect on average to a stationer system. In regards to superposition of states, our model supports the interaction of Schr\"{o}dinger equation introduced by Bohm that at a definite time and position the particle only occupies one state but, the equation should cover all of the possible states of the particle.

	\begin{acknowledgments}
	
The work\,\, was\,\, supported\,\, by \,\,grand\,\, Penelitian\,\, Dosen\,\, Muda \,\,in \,\,2011 \,\,(contract \,\,no. \,\,LPPM-UGM /1506/BID.I/201), and I am indebted to Prof. Andrew Strominger, Wa Ode Kamaria and Sylvia Hase for their great supporting.

	\end{acknowledgments}
	
	\bibliographystyle{plain}  

 \end{document}